\documentclass[12pt]{article}
\usepackage{amsthm, amsmath, natbib}
\usepackage{enumerate}
\usepackage{psfrag,epsf}
\usepackage{graphicx}
\usepackage{multirow}
\usepackage{subcaption}
\usepackage{mathtools}
\usepackage{url} 
\usepackage{titlesec}
\usepackage{hhline}
\titlelabel{\thetitle.\quad}
\usepackage{titlesec}
\usepackage{titletoc}
\usepackage[title,titletoc]{appendix}

\usepackage[doublespacing]{setspace}
\usepackage{tocloft}
\usepackage{color}
\usepackage{caption}

\newtheorem*{thm*}{Theorem}

\newtheorem*{prop*}{Proposition}

\newcommand{\blind}{0}

\begin{document}
\date{}

\newcommand{\ssection}[1]{\section[#1]{\centering #1}}

\def\spacingset#1{\renewcommand{\baselinestretch}%
{#1}\small\normalsize} \spacingset{1}

\if0\blind
{
  \title{\bf  Data transforming augmentation for heteroscedastic models}%
  \author{Hyungsuk Tak$^{\dag}$, Kisung You$^{\S}$, Sujit \textcolor{black}{K.}~Ghosh$^{\P}$, Bingyue Su$^{\S}$, and Joseph Kelly$^{\ddag}$\thanks{Corresponding author: josephkelly@google.com} \vspace{.4cm}\\
     $^{\dag}$Pennsylvania State University, $^{\S}$University of Notre Dame,\\ $^{\P}$North Carolina State University, $^{\ddag}$Google LLC}
  \maketitle
} \fi

\if1\blind
{
   \title{\bf Data transforming augmentation for heteroscedastic models}%
   \author{~}
   \maketitle
} \fi

\begin{abstract}
\noindent Data augmentation (DA) turns seemingly intractable computational problems into simple ones by augmenting latent missing data. In addition to  computational simplicity, it is now well-established that DA equipped with a deterministic transformation can improve the convergence speed of iterative algorithms such as an EM algorithm or Gibbs sampler. In this article, we outline a  framework for the transformation-based DA, which we call data transforming augmentation (DTA), allowing augmented data to be a deterministic function of latent and observed data, and unknown parameters. Under this framework, we  investigate a novel DTA scheme that turns heteroscedastic models into  homoscedastic ones to take advantage of simpler computations typically available in homoscedastic cases. Applying this DTA scheme to fitting linear mixed models, we demonstrate simpler computations and faster convergence rates of resulting iterative algorithms, compared with those  under a non-transformation-based DA scheme. We also fit a Beta-Binomial model using the proposed DTA scheme, which enables sampling approximate marginal posterior distributions  that are available only under homoscedasticity. An R package \texttt{Rdta} is publicly available at CRAN.
\end{abstract}

\noindent%

\spacingset{1.1} 

\newpage

\section{Introduction}\label{sec1}

Data augmentation (DA) is an art to convert a complicated statistical computation into a simpler one \citep{tanner1987calculation, van2001art}. Its key idea is to introduce latent missing data $y^{\textrm{mis}}$ to form augmented data $y^{\textrm{aug}}=(y^{\textrm{obs}}, y^{\textrm{mis}})$ in a way that  an augmented data likelihood function of an unknown parameter vector $\theta$, i.e.,
$$
L(\theta; y^{\textrm{aug}})\propto f(y^{\textrm{aug}}\mid\theta),
$$ 
is easier to handle than the observed data likelihood function, 
$$
L(\theta; y^{\textrm{obs}})\propto \int f(y^{\textrm{aug}} \mid \theta)~dy^{\textrm{mis}}=f(y^{\textrm{obs}}\mid\theta).
$$ (All density functions in this article are defined with respect to a common dominating measure, e.g., Lebesgue or counting measure.)

Although computational simplicity can be achieved via DA, it does not necessarily lead to faster convergence rates of iterative algorithms  \citep{meng1997folksong}; we consider an EM algorithm \citep{dempster1977maximum} as a deterministic iterative algorithm  and a Gibbs sampler \citep{geman1984stochastic} as  a stochastic counterpart. As a possible remedy, researchers have demonstrated that certain DA schemes equipped with  deterministic  transformations, i.e., data transforming augmentation (DTA), can facilitate the resulting statistical computations and improve convergence rates of iterative algorithms as well. Such efforts include (but are not limited to) the  alternating expectation conditional maximization \textcolor{black}{algorithm} \citep{meng1997folksong}, conditional and marginal data augmentation schemes \citep{meng1999seeking}, re-parametrization of augmentation scheme \citep{papa2007repara, papa2008repara}, and ancillarity-sufficiency interweaving strategy \citep{yu2011}. However, their deterministic  transformations  are limited to those of missing data $y^{\textrm{mis}}$ and model parameters~$\theta$,  although DTA broadly allows transformations with observed data $y^{\textrm{obs}}$ as well.

To emphasize such extensive applicability of DTA, we use a  framework that allows the augmented data $y^{\textrm{aug}}$ to be a deterministic function of both $y^{\textrm{mis}}$ and $y^{\textrm{obs}}$, which may also depend on $\theta$. For example, $y^{\textrm{aug}}=h(y^{\textrm{obs}},  y^{\textrm{mis}})$ for some bijection mapping $h$ \citep{vandyk2010fertilizing} or possibly in a parameter-dependent form, i.e., $y^{\textrm{aug}}=h(y^{\textrm{obs}},  y^{\textrm{mis}}; \theta)$ \citep{papa2007repara}. If $h$ is an identity function,  then $y^{\textrm{aug}}=(y^{\textrm{obs}},  y^{\textrm{mis}})$, and DTA becomes non-transformation-based DA (which we simply call DA hereafter).  Therefore, the key to DTA is to choose a deterministic function~$h$ that results in either simpler computations or faster convergence rates of iterative algorithms.

 \textcolor{black}{This aspect of choosing $h$ is in line with DA; in practice it may be challenging to find a useful transformation $h$  as it is the case for DA in finding a useful augmentation scheme with appropriate missing data $y^{\textrm{mis}}$. But, at the same time, DTA can provide more possibilities to improve computational efficiency because  even if there is no obvious DA solution, one may still be able to try various deterministic transformations for desired computational advantages. For example, to improve computational efficiency, researchers may already know which properties are desired for the augmented data, and this knowledge may in turn enable finding  appropriate missing data and transformations that lead to the desired augmented data.}


\textcolor{black}{In this paper}, we investigate a novel DTA scheme to convert \textcolor{black}{a heteroscedastic model to a homoscedastic one} via a deterministic transformation of both $y^{\textrm{mis}}$ and $y^{\textrm{obs}}$, i.e., when a model for $y^{\textrm{obs}}$ is heteroscedastic, a model for $y^{\textrm{aug}}=h(y^{\textrm{obs}},  y^{\textrm{mis}})$ is homoscedastic. \textcolor{black}{Here, the missing data $y^{\textrm{mis}}$ and deterministic transformation $h$ are carefully chosen to achieve the conversion from heteroscedasticity to homoscedasticity.  Computational simplicity that is typically available under homoscedasticity is one motivation of the proposed transformation. Computational and inferential issues associated with heteroscedastic models have been discussed in the literature; see \cite{everson2000inference} for linear mixed models, \cite{staudenmayer2008density} for density estimation, and \cite{xianchao2012sure} for hierarchical models. We also investigate whether such computational simplicity results in faster convergence rate of iterative algorithms.}  

\textcolor{black}{We illustrate our idea with two commonly encountered examples of heteroscedastic models. First,} we apply this DTA scheme to fitting linear mixed models, which is first introduced for a univariate case in \cite{kelly2014advances}. We provide a theoretical justification for the result of \cite{kelly2014advances} about the improvement of convergence rates of \textcolor{black}{DTA-based} iterative algorithms, and generalize his univariate case to a multivariate one.  Additionally, as a part of numerical illustrations, we conduct  a simulation study and analyze realistic hospital profiling data to demonstrate faster convergence rates of iterative algorithms under DTA. \textcolor{black}{Secondly}, we apply the proposed DTA scheme to fitting a Beta-Binomial model \textcolor{black}{on over-dispersed Binomial data with heterogeneous numbers of trials. Our motivation is that} approximate marginal posterior distributions are available in a closed form \textcolor{black}{if the numbers of trials are homogeneous}  \citep{everson2002betabin}. Transforming \textcolor{black}{heteroscedasticity} to homoscedasticity, the DTA scheme enables sampling the approximate  posterior distributions even in a heteroscedastic case. We numerically illustrate the approximation accuracy  \textcolor{black}{by analyzing realistic baseball data}.

\textcolor{black}{In what follows, Section~\ref{sec2} specifies the DTA framework, Section~\ref{sec3} describes the applications of the proposed DTA scheme to linear mixed  and Beta-Binomial models, and finally Section~\ref{sec4} shows our future directions.} An R package, \texttt{Rdta},  to fit univariate and multivariate linear mixed models via DA- and DTA-based iterative algorithms is publicly available on CRAN\footnote{https://cran.r-project.org/package=Rdta}.

\section{Data transforming augmentation}\label{sec2}

Recall that under a DA scheme \citep{tanner1987calculation}, a Gibbs-type algorithm iteratively samples the following  two conditional distributions: 
\begin{equation}\label{da}
[y^{\textrm{mis}}\mid y^{\textrm{obs}}, \theta]~~~\textrm{and}~~~[\theta\mid y^{\textrm{aug}}],
\end{equation} 
where the notation $[a\mid b]$ denotes the conditional distribution of $a$ given $b$, the augmented data $y^{\textrm{aug}}$ are $(y^{\textrm{obs}}, y^{\textrm{mis}})$, and \textcolor{black}{the posterior} density functions \textcolor{black}{of~\eqref{da}} are proportional to $L(\theta; y^{\textrm{aug}})$ multiplied by a prior $\pi(\theta)$.  This Gibbs sampler marginally preserves the target posterior  distribution whose density function is $p(\theta\mid y^{\textrm{obs}})\propto L(\theta; y^{\textrm{obs}})\pi(\theta)$.


DTA inserts a deterministic  transformation step in the middle of~\eqref{da}. The resulting Gibbs-type algorithm with a parameter-free transformation, $y^{\textrm{aug}}=h(y^{\textrm{mis}}, y^{\textrm{obs}})$,  iteratively samples the following three conditional distributions:
\begin{equation}\label{dta1}
[y^{\textrm{mis}}\mid y^{\textrm{obs}}, \theta],~~~  [y^{\textrm{aug}}\mid y^{\textrm{obs}}, y^{\textrm{mis}}]~~~\textrm{and}~~~  [\theta\mid y^{\textrm{aug}}],
\end{equation}
where the density function of $[y^{\textrm{aug}}\mid y^{\textrm{obs}}, y^{\textrm{mis}}]$ in the middle is one if $y^{\textrm{aug}}=h(y^{\textrm{mis}}, y^{\textrm{obs}})$ and zero otherwise.   Since the middle step in~\eqref{dta1} is  a deterministic transformation, sampling the three conditional distributions in~\eqref{dta1} is essentially the same as sampling the following two conditional distributions:
\begin{equation}\label{dta2}
[y^{\textrm{aug}}\mid y^{\textrm{obs}}, \theta]~~~\textrm{and}~~~[\theta\mid y^{\textrm{aug}}].  
\end{equation}
Also, since  we can sample the first conditional distribution  in~\eqref{dta2}, i.e., $[y^{\textrm{aug}}\mid y^{\textrm{obs}}, \theta]$, by  sampling $[y^{\textrm{mis}}\mid y^{\textrm{obs}}, \theta]$ and $[y^{\textrm{aug}}\mid y^{\textrm{obs}}, y^{\textrm{mis}}, \theta]$ in turn, a Gibbs-type algorithm for a parameter-dependent version of DTA sequentially samples
\begin{equation}\label{dta1p}
[y^{\textrm{mis}}\mid y^{\textrm{obs}}, \theta],~~~  [y^{\textrm{aug}}\mid y^{\textrm{obs}}, y^{\textrm{mis}}, \theta]~~~\textrm{and}~~~  [\theta\mid y^{\textrm{aug}}],
\end{equation}
where the density function of $[y^{\textrm{aug}}\mid y^{\textrm{obs}}, y^{\textrm{mis}}, \theta]$ in the middle is one if $y^{\textrm{aug}}=h(y^{\textrm{mis}}, y^{\textrm{obs}}; \theta)$ and zero otherwise. This shows that both parameter-free and parameter-dependent  DTA schemes in~\eqref{dta1} and~\eqref{dta1p} are marginally equivalent to~\eqref{dta2}. 

Many researchers have \textcolor{black}{established} that such DTA schemes can improve the convergence speed of iterative algorithms in the literature. For example, conditional and marginal DA schemes \citep{meng1997folksong, meng1999seeking}, i.e., working-parameter-based approaches to  efficient DA scheme\textcolor{black}{s}, are the parameter-free and  parameter-dependent DTA schemes in~\eqref{dta1} and~\eqref{dta1p}, respectively. This is because the conditional DA  involves a deterministic transformation of missing data with a  working parameter $\alpha$ that is fixed at a constant; for instance, $y^{\textrm{aug}}=h(y^{\textrm{obs}}, y^{\textrm{mis}})=(y^{\textrm{obs}}, \alpha y^{\textrm{mis}})$. Unlike the conditional DA, the marginal DA treats the working parameter $\alpha$ as \textcolor{black}{an} unknown parameter and marginalizes it via a Gibbs-type implementation. Under the marginal DA the same transformation  is considered as a parameter-dependent form, i.e.,  $y^{\textrm{aug}}=h(y^{\textrm{obs}}, y^{\textrm{mis}}; \theta)=(y^{\textrm{obs}}, \alpha y^{\textrm{mis}})$, where $\alpha\in \theta$. A parameter-expanded EM algorithm \citep{liu1998pxem} and  a re-parametrization of an augmentation scheme \citep{papa2007repara} also adopt such a parameter-dependent data transformation of the latent missing data $y^{\textrm{mis}}$ and parameters $\theta$. 


We can  extend the DTA framework, considering that sampling  $[y^{\textrm{aug}}\mid y^{\textrm{obs}}, y^{\textrm{mis}}]$ in \textcolor{black}{the middle of}~\eqref{dta1} is equivalent to  sampling $[\theta \mid y^{\textrm{obs}}, y^{\textrm{mis}}]$ and $[y^{\textrm{aug}}\mid y^{\textrm{obs}}, y^{\textrm{mis}}, \theta]$ in a sequence. In this case, \eqref{dta1} becomes
\begin{equation}\label{dta3}
[y^{\textrm{mis}}\mid y^{\textrm{obs}}, \theta],~~~  [\theta \mid y^{\textrm{obs}}, y^{\textrm{mis}}],~~~ [y^{\textrm{aug}}\mid y^{\textrm{obs}}, y^{\textrm{mis}}, \theta]~~~\textrm{and}~~~  [\theta\mid y^{\textrm{aug}}],
\end{equation}
where the first two conditional distributions are the same as~\eqref{da} under DA, and the density of $[y^{\textrm{aug}}\mid y^{\textrm{obs}}, y^{\textrm{mis}}, \theta]$ in the third step is one if $y^{\textrm{aug}}=h(y^{\textrm{obs}}, y^{\textrm{mis}}; \theta)$ and zero otherwise. A global interweaving strategy of an ancillarity-sufficiency interweaving strategy \citep{yu2011} adopts this extended version of parameter-dependent DTA. In the global interweaving strategy, the intermediate update of $\theta$ in the second step of~\eqref{dta3} is achieved under an ancillary parameterization (also called centered parametrization in \cite{papa2007repara}). The  deterministic transformation in the third step of~\eqref{dta3} turns the ancillary parameterization to a sufficient parameterization (also called non-centered parametrization in \cite{papa2007repara}). Thus, the final update of $\theta$ is achieved under the sufficient parameterization. 



\cite{yu2011} have already \textcolor{black}{deduced} that \textcolor{black}{the global interweaving strategy (DTA in~\eqref{dta3})} marginally preserves the target posterior distribution, $p(\theta\mid y^{\textrm{obs}})$. \textcolor{black}{However, their argument can be used for a more general purpose, i.e., not only for proving the correct marginal posterior distribution of the global interweaving strategy, but also for proving that of any DTA scheme. For this purpose, we \textcolor{black}{re-state their proof in a more general context (for any deterministic transformations), using} the DTA scheme in~\eqref{dta2} because} the other versions of DTA in~\eqref{dta1}, \eqref{dta1p}, and \eqref{dta3} are marginally equivalent to \textcolor{black}{DTA in}~\eqref{dta2}. The transition kernel density of DTA in~\eqref{dta2} is
\begin{equation}\label{kernel}
K(\theta\mid\theta^\ast)=\int p_2(\theta\mid y^{\textrm{aug}})  p_1(y^{\textrm{aug}}\mid y^{\textrm{obs}}, \theta^\ast)~dy^{\textrm{aug}},
\end{equation}
where $\theta^\ast$ denotes the sampled value of $\theta$ at the previous iteration, and each subscript of  density functions in the integrand indicates the sampling sequence. This kernel density preserves the target  distribution $p(\theta\mid y^{\textrm{obs}})$ as follows:
\begin{align}
\begin{aligned}\label{stationary}
\int K(\theta\mid\theta^\ast)p(\theta^\ast\mid y^{\textrm{obs}})~d\theta^\ast&=\int\!\!\int p_2(\theta\mid y^{\textrm{aug}})  p_1(y^{\textrm{aug}}\mid y^{\textrm{obs}}, \theta^\ast)p(\theta^\ast\mid y^{\textrm{obs}})~d\theta^\ast dy^{\textrm{aug}}\\
&=\int p_2(\theta\mid y^{\textrm{aug}})  p(y^{\textrm{aug}}\mid y^{\textrm{obs}})~dy^{\textrm{aug}}=p(\theta\mid y^{\textrm{obs}}).
\end{aligned}
\end{align}
Therefore, all of the parameter-free, parameter-dependent, and parameter-extended versions of DTA  preserve the same marginal posterior distribution $p(\theta\mid y^{\textrm{obs}})$, as DA does. 


If a Gibbs sampler is a stochastic implementation of DTA, its deterministic counterpart is an EM algorithm \citep{dempster1977maximum}. The $Q$ function in the E-step of an EM algorithm under either DTA or DA is
\begin{equation}\label{qfunc}
Q(\theta\mid \theta^\ast)=E(\log(f(y^{\textrm{aug}}\mid \theta))\mid y^{\textrm{obs}}, \theta^\ast),
\end{equation}
where \textcolor{black}{under DA $y^{\textrm{aug}}=(y^{\textrm{obs}}, y^{\textrm{mis}})$, and under DTA  $y^{\textrm{aug}}=h(y^{\textrm{obs}}, y^{\textrm{mis}})$ or $y^{\textrm{aug}}=h(y^{\textrm{obs}}, y^{\textrm{mis}}; \theta)$  for respective  parameter-free and parameter-dependent DTA}. As for DTA, deriving the density function of $[y^{\textrm{aug}}\mid \theta]$ may involve a \textcolor{black}{Jacobian} term \citep[Equation~2.6,][]{meng1999seeking}. Since the only difference between DTA and DA is  the conditional distribution of the augmented data $[y^{\textrm{aug}}\mid\theta]$ \textcolor{black}{with} the same marginal $[y^{\textrm{obs}}\mid\theta]$,  an EM algorithm based on DTA can be regarded as an EM algorithm under a different DA scheme. Thus, \textcolor{black}{both DA-based and DTA-based} EM algorithms share the same monotonic hill-climbing property, i.e.,  monotonically increasing the likelihood at each iteration \citep[Section~3,][]{dempster1977maximum}. 


\textcolor{black}{We note that within  Gibbs- and EM-type iterative algorithms, the deterministic transformation $y^{\textrm{aug}}=h(y^{\textrm{obs}}, y^{\textrm{mis}})$ is treated as a one-to-one function  between $y^{\textrm{aug}}$ and $y^{\textrm{mis}}$ because  $y^{\textrm{obs}}$ is  fixed at a constant in the iterative algorithms and $h$ is known in advance. (This also holds for parameter-dependent DTA because both $y^{\textrm{obs}}$ and $\theta$ are given when $y^{\textrm{aug}}$ is updated  in~\eqref{dta2} or marginalized in~\eqref{qfunc}.)  Thus, the  scale of DA is seamlessly transferred to that of DTA; an EM-type algorithm  accounts for a scale change via a Jacobian term and deterministic transformations are straightforward within a Gibbs-type algorithm, e.g., DTA in~\eqref{dta1}. After all, since both DA and DTA models  preserve the same marginal model, any inferences including predictions and effects of covariates  would be identical under both DA and DTA.}

\section{Fitting linear mixed models via DTA}\label{sec3}

\subsection{A univariate case}\label{sec31}

Let us assume that $y^{\textrm{obs}}_i$ is an unbiased estimate of random effect $\theta_i$ with known (or accurately estimated) measurement error variance $V_i$ of group~$i$ ($i=1, 2, \ldots, k$). We specify a univariate linear mixed model \citep{efron1975data, kass1989approx, daniels1999prior, morris2012shrinkage} as follows:
\begin{equation}\label{equ1}
y^{\textrm{obs}}_i\mid\theta_i\stackrel{\textrm{ind.}}{\sim} \textrm{N}_1(\theta_i,~ V_i)~~~\textrm{and}~~~\theta_i\mid A, \beta \stackrel{\textrm{ind.}}{\sim} \textrm{N}_1(\boldsymbol{x}_i^{\top}\boldsymbol{\beta},~ A),
\end{equation}
where $\boldsymbol{x}_i$ is a covariate vector of length $m$, and both  regression coefficients $\boldsymbol{\beta}=(\beta_1, \ldots, \beta_m)^{\top}$ and variance component $A$ of the Gaussian prior distribution of random effects are  unknown  parameters of interest. One way to infer the parameters of interest is to marginalize the random effects and use the  observed data likelihood, $L(A, \boldsymbol{\beta}; y^{\textrm{obs}})=f(y^{\textrm{obs}}\mid A, \boldsymbol{\beta})$, where $[y^{\textrm{obs}}_i\mid A, \boldsymbol{\beta}]\sim \textrm{N}_1(\boldsymbol{x}^{\top}_i\boldsymbol{\beta},~ A+V_i)$ independently.  For a  Bayesian inference, \cite{kelly2014advances} adopts Stein's harmonic prior \textcolor{black}{\citep{morris2012shrinkage}}, i.e., $p(A, \boldsymbol{\beta})\propto I_{A>0}$, for good frequency properties  and proves that the resulting posterior is proper if $k\ge m+3$. 

It is possible to  sample $[A\mid y^{\textrm{obs}}]$ directly via an inverse-Gamma distribution in a homoscedastic case ($V_i=V$ for all $i$), while it is not possible in a heteroscedastic case \citep{everson2000inference}. This is the motivation for \cite{kelly2014advances}   to introduce missing data $y^{\textrm{mis}}_i$ in a  way that a convex combination of  $y^{\textrm{mis}}_i$  and $y^{\textrm{obs}}_i$ becomes homoscedastic, i.e., 
\begin{equation}\label{equ2}
y^{\textrm{aug}}_i=h(y^{\textrm{obs}}_i, y^{\textrm{mis}}_i)=(1-w_i)y^{\textrm{obs}}_i+w_iy^{\textrm{mis}}_i,
\end{equation}
where $w_i=1-V_\textrm{min} /V_i$ is a weight of the convex combination, $V_\textrm{min}=\min(V_1, V_2, \ldots, V_k)$ is the minimum variance, and $[y^{\textrm{mis}}_i\mid \theta_i]\sim \textrm{N}_1(\theta_i, w_i^{-1}V_{\textrm{min}})$  with an assumption that $y^{\textrm{mis}}_i$ is conditionally independent of $y^{\textrm{obs}}_i$ given $\theta_i$.  \textcolor{black}{(Dr.~Carl N.~Morris at Harvard University, the  supervisor of \cite{kelly2014advances}, is also credited for the development of this DTA scheme.)}

\textcolor{black}{The augmented data are homoscedastic, i.e., $[y^{\textrm{aug}}_i\mid \theta_i]\sim \textrm{N}_1(\theta_i,~ V_\textrm{min})$  for all $i$. This is because if the measurement error variance of group~$i$ is the same as the minimum variance (i.e., $V_i=V_{\textrm{min}}$), then the corresponding weight  $w_i$ becomes 0, and thus $y^{\textrm{aug}}_i$  in~\eqref{equ2} is set to $y^{\textrm{obs}}_i$. As a result, $[y^{\textrm{aug}}_i\mid \theta_i]$ is the same as $[y^{\textrm{obs}}_i\mid \theta_i]\sim \textrm{N}_{1}(\theta_i,~ V_{\textrm{min}})$ for group~$i$. Also, if  $V_i\neq V_{\textrm{min}}$, the mean and variance of $y^{\textrm{aug}}_i$ given $\theta_i$ are still $\theta_i$ and $V_{\textrm{min}}$, respectively, because}
\begin{align}
\textcolor{black}{E(y^{\textrm{aug}}_i\mid \theta_i)}~&\textcolor{black}{=(1-w_i)E(y^{\textrm{obs}}_i\mid\theta_i)+w_iE(y^{\textrm{mis}}_i\mid \theta_i)=(1-w_i)\theta_i+w_i\theta_i=\theta_i,}\nonumber\\
\textcolor{black}{\textrm{Var}(y^{\textrm{aug}}_i\mid \theta_i)}~&\textcolor{black}{=(1-w_i)^2\textrm{Var}(y^{\textrm{obs}}_i\mid \theta_i)+w_i^2\textrm{Var}(y^{\textrm{mis}}_i\mid \theta_i)}\nonumber\\
&\textcolor{black}{=(1-w_i)^2V_i+w_iV_{\textrm{min}}=V_{\textrm{min}}^2/V_i+(V_{\textrm{min}}-V_{\textrm{min}}^2/V_i)=V_{\textrm{min}}}.\nonumber
\end{align}


\cite{kelly2014advances} derives both DTA- and DA-based Gibbs samplers to sample the full posterior $p(A, \boldsymbol{\beta}\mid y^{\textrm{obs}})\propto L(A, \boldsymbol{\beta}; y^{\textrm{obs}})I_{A>0}$ and corresponding EM algorithms to find the posterior modes (or maximum likelihood estimates) of $A$ and $\beta$. The DA-based approach treats random effects as missing data, i.e., $y^{\textrm{aug}}=(y^{\textrm{obs}}, \theta)$, which is commonly adopted in the literature \citep{morris1987comment}. We specify  details of these Gibbs samplers and EM algorithms in the supplementary material (Section~A). 

To quickly illustrate the convergence speed of DTA- and DA-based iterative algorithms in a univariate case, we simulate 50 observations using~\eqref{equ1} without covariate information (i.e., $\beta=\beta_1$ and $x_i=1$ for all $i$). We first generate $\theta_i$'s given $\beta=0$ and $A=5$, randomly draw  $V_i$'s  from $\textrm{N}_1(10, 2^2)$ for severe heteroscedasticity, and then generate $y^{\textrm{obs}}$ conditioning on  the sampled $\theta_i$'s and $V_i$'s.

We implement Gibbs samplers and EM algorithms derived under both DTA and DA schemes. For each Gibbs sampler we draw 510,000 posterior samples of $A$ and $\beta$, where the initial value of $\beta$ is randomly generated from $\textrm{N}_1(0, 1)$ and that of $A$ is randomly set to one of the known measurement variances, $V_i$'s. It takes 14.48   seconds  for the DTA-based Gibbs sampler and 11.43 seconds for the DA-based one \textcolor{black}{to obtain a single Markov chain of length 510,000 without parallelization; the} CPU time is obtained from a desktop equipped with 4-core Intel i7-processor at 3.5 GHz and 32 GB of memory. We discard the first 10,000 samples as burn-in. To implement the corresponding EM algorithms, we fix the initial values of $(\beta, A)$ at $(0, 1)$. With a tolerance level set to $10^{-10}$, it takes 0.138 second (15 iterations) for the DTA-based EM algorithm, and 0.140 second (89 iterations) for the DA-based one. 

\begin{figure}[t!]
\begin{center}
\includegraphics[scale=0.348]{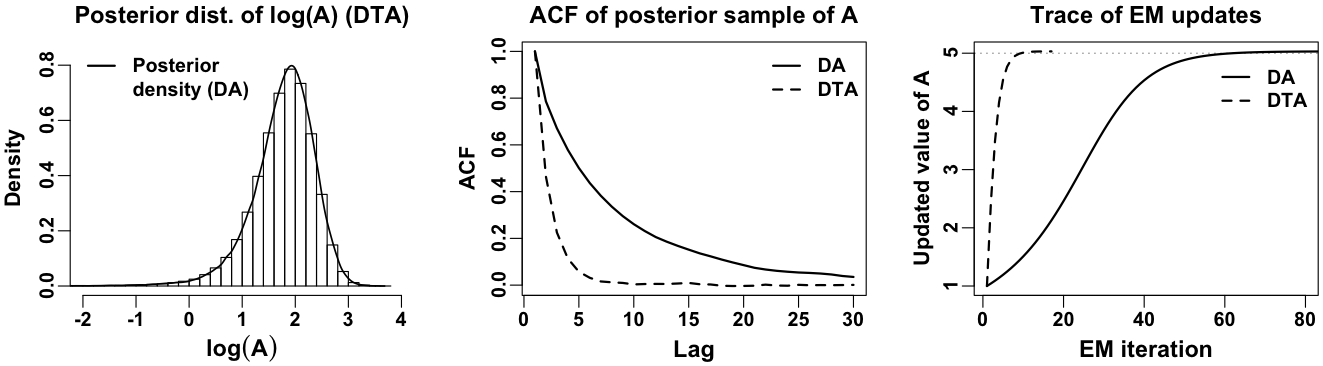}
\caption{The  results of Gibbs sampling and EM mound-climbing. The first panel shows the posterior distribution of $\log(A)$ obtained under DTA, and the posterior density of $\log(A)$ obtained under DA is superimposed to show their converged stationary distribution. In the second panel, the auto-correlation function of the posterior sample of $A$ under DTA (dashed curve) decreases faster than that under   DA (solid curve). The third panel  exhibits that the mode-climbing of EM algorithm  is faster under DTA  (dashed curve).}
\label{fig1}
\end{center}
\end{figure}

Figure~\ref{fig1} displays  outcomes of fitting the model with DTA- and DA-based iterative algorithms. The first panel displays the posterior distribution of $\log(A)$ obtained with DTA.  We also superimpose the posterior density of $\log(A)$ (solid curve) obtained with DA to confirm that their stationary distributions are consistent; for this purpose we use \texttt{density}, a built-in function in R \citep{r2019}.  In the second panel, we use a dashed curve to denote the auto-correlation function of the  posterior sample of $A$ obtained with  DTA,  and a solid curve to represent the one obtained with DA.  The auto-correlation function under DTA decreases more quickly than that under DA does.  The effective sample size per CPU time is 13,778 under DTA and 3,104 under DA, indicating that the former is about 4.4 times larger than the latter.   The last panel compares the update history of $A$ at \textcolor{black}{each} EM iteration under both augmentation schemes. The updated values under DTA (denoted by the dashed curve) approach  the maximum likelihood estimate of $A$ more quickly than those under DA (the solid curve) with almost identical CPU times.

In this univariate case, we can also use the matrix  rate of convergence, or so-called matrix fraction of missing information, to compare  convergence rates of EM algorithms \citep{dempster1977maximum, vandyk2010fertilizing}.  The largest eigenvalue of this  matrix rate represents the global convergence rate of an EM algorithm; the larger the eigenvalue, the slower the EM algorithm.  It is worth noting that this global convergence rate of an EM algorithm can approximate the geometric convergence rate of a Gibbs-type algorithm  in practice \citep[Section~2.2,][]{meng1999seeking}, \textcolor{black}{i.e.,} a Gibbs-type algorithm may converge fast if the corresponding EM algorithm does. 

Using the matrix rate of convergence, we provide both theoretical derivation and numerical illustration to show superior performance of the DTA scheme over the DA scheme. In \textcolor{black}{the} Appendix, we prove that  for the linear mixed model in~\eqref{equ1}, the matrix rate under DA subtracted by that under DTA is positive definite on average. This means that the convergence rate of the EM algorithm under DTA is expected to be faster than that under DA. In our numerical study, the largest eigenvalue of the matrix  rate  under DTA is 0.489 and that under DA is 0.889. This clearly indicates that the convergence rate of the EM algorithm under DTA is faster than that under DA. \textcolor{black}{The sum} of these theoretical and empirical evidence corroborates the previous comparison result between the Gibbs samplers derived under both augmentation schemes.

\subsection{A multivariate case}\label{sec32}

We generalize the DTA scheme of \cite{kelly2014advances} to a multivariate case. The linear mixed model for $p$-variate observations and random effects \citep{everson2000inference, gasparrini2012mlmm}  is
\begin{equation}\label{mlmm}
[\boldsymbol{y}_i^{\textrm{obs}}\mid \boldsymbol{\theta}_i]  \stackrel{\textrm{ind.}}{\sim} \textrm{N}_p(\boldsymbol{\theta}_i,~ \boldsymbol{V}\!_i)~~~\textrm{and}~~~[\boldsymbol{\theta}_i\mid \boldsymbol{A}, \boldsymbol{\beta}] \stackrel{\textrm{ind.}}{\sim}  \textrm{N}_p(X_i \boldsymbol{\beta},~ \boldsymbol{A}),
\end{equation}
where $X_i=I_p \otimes \boldsymbol{x}_i^{\top}$ is a $p$ by $mp$ block diagonal covariate matrix defined as a \textcolor{black}{K}ronecker product of a $p$ dimensional identity matrix $I_p$  and a row vector of $m$ covariates (i.e., $\boldsymbol{x}_i^{\top}=\{x_{i1}, x_{i2}, \ldots, x_{im}\}$ appears along the diagonal), $\boldsymbol{\beta}=\{\beta_1, \beta_2, \ldots, \beta_{mp}\}$ is a vector of length $mp$  for unknown regression coefficients, $\boldsymbol{V}\!_i$ is a known positive definite matrix, and $\boldsymbol{A}$ is an unknown $p$ by $p$ covariance matrix of random effects. The marginal distribution of $\boldsymbol{y}_i^{\textrm{obs}}$  is $[\boldsymbol{y}_i^{\textrm{obs}}\mid \boldsymbol{A}, \boldsymbol{\beta}]\sim \textrm{N}_p(X_i \boldsymbol{\beta},~ \boldsymbol{A}+\boldsymbol{V}\!_i)$,  and the corresponding observed data likelihood function is $L(\boldsymbol{A}, \boldsymbol{\beta}; \boldsymbol{y}^{\textrm{obs}})=f(\boldsymbol{y}^{\textrm{obs}}\mid \boldsymbol{A}, \boldsymbol{\beta})$. For a Bayesian inference, we use a multivariate version of Stein's harmonic prior, $p(\boldsymbol{A}, \boldsymbol{\beta})\propto I_{\mid \boldsymbol{A}\mid>0}$, which is known to have good frequency properties with guaranteed posterior propriety when $k\ge m+p+2$ \citep{everson2000inference, tak2017frequency}.

Even though $\boldsymbol{y}_i^{\textrm{obs}}$ is heteroscedastic, the following augmented data are homoscedastic:
\begin{equation}\label{multivariatedata transforming augmentation}
\boldsymbol{y}_i^{\textrm{aug}}=h(\boldsymbol{y}_i^{\textrm{obs}}, \boldsymbol{y}_i^{\textrm{mis}})=(I_p - W_i) \boldsymbol{y}_i^{\textrm{obs}}+W_i\boldsymbol{y}_i^{\textrm{mis}},
\end{equation}
where $W_i$ is defined as  $I_p - \boldsymbol{V}_{\textrm{min}}^{0.5}\boldsymbol{V}^{-1}_i\boldsymbol{V}_{\textrm{min}}^{0.5}$  with $\boldsymbol{V}_{\textrm{min}}^{0.5}$ denoting the symmetric matrix square root of a positive definite matrix $\boldsymbol{V}\!_{\textrm{min}}$, and 
\begin{equation}\label{covmatrix}
[\boldsymbol{y}_{i}^{\textrm{mis}}\mid \boldsymbol{\theta}_i]\sim \textrm{N}_p(\boldsymbol{\theta}_i,~ \boldsymbol{V}_{\textrm{min}}^{0.5}W_i^{-1}\boldsymbol{V}_{\textrm{min}}^{0.5}).
\end{equation}
We assume that $\boldsymbol{y}_{i}^{\textrm{mis}}$ and $\boldsymbol{y}_{i}^{\textrm{obs}}$ are conditionally independent given $\boldsymbol{\theta}_i$. Since $\boldsymbol{V}\!_i$ can be decomposed into $Q_i\Lambda_iQ_i^{\top}$, where $Q_i$ is an orthogonal matrix and $\Lambda_i$ is a diagonal matrix whose diagonal elements are $(\lambda_{1i}, \lambda_{2i}, \ldots, \lambda_{pi})^{\top}$, we set $\boldsymbol{V}\!_{\textrm{min}}=\lambda_{\min}I_p$, where $\lambda_{\min}$ is defined as the minimum eigenvalue among all $pk$ eigenvalues, $\lambda_{ji}$'s ($j=1, \ldots, p; i=1, \ldots, k$). Then, this augmentation scheme reduces to the univariate case of \cite{kelly2014advances} if $p=1$.  

Unlike the univariate case, $W_i$ becomes singular if $\lambda_{\min}=\lambda_{ji}$ for some~$j$ of group $i$, and thus the inverse of $W_i$ in~\eqref{covmatrix} does not exist. This does not cause any problem in posterior inference because the resulting conditional posterior distribution of $\boldsymbol{y}_{i}^{\textrm{mis}}$ does not involve the inverse of $W_i$; see (23) in the supplementary material. Alternatively, we can define $\boldsymbol{V}\!_{\textrm{min}}=(0.999\times \lambda_{\min})I_p$ to guarantee the non-singularity of $W_i$; the resulting posterior inference will be almost identical to the one with $\boldsymbol{V}\!_{\textrm{min}}=\lambda_{\min}I_p$.

Since the augmented data are homoscedastic, i.e.,  $[\boldsymbol{y}_i^{\textrm{aug}}\mid \boldsymbol{\theta}_i]  \sim \textrm{N}_p(\boldsymbol{\theta}_i, \boldsymbol{V}_{\textrm{min}})$, we can directly sample $[\boldsymbol{A}, \boldsymbol{\beta}\mid  \boldsymbol{y}^{\textrm{aug}}]$ from standard family distributions \citep{everson2000inference}, as is the case in the univariate case. Then the Gibbs  sampler under DTA iteratively samples $[\boldsymbol{y}^{\textrm{aug}}\mid \boldsymbol{y}^{\textrm{obs}}, \boldsymbol{A}, \boldsymbol{\beta}]$ and $[\boldsymbol{A}, \boldsymbol{\beta}\mid  \boldsymbol{y}^{\textrm{aug}}]$, where $[\boldsymbol{y}^{\textrm{aug}}\mid \boldsymbol{y}^{\textrm{obs}}, \boldsymbol{A}, \boldsymbol{\beta}]$ is multivariate Gaussian. In the supplementary material (Section~B), we specify details of  the Gibbs sampler and corresponding EM algorithm based on DTA and those based on typical DA.

\begin{table}[b!]
\caption{The  hospital profiling data  summarize the interviews with 1,869 patients, showing for hospital $i$ the percentage of non-surgical issues ($y_{1i}$) and that of surgical issues ($y_{2i}$),  a severity measure ($x_i$), and the number of patients ($n_i$).}
{\small\begin{tabular}{ccccccccccccccc}
$i$ & $y_{1i}$ & $y_{2i}$ & $x_{i}$  & $n_i$ &  ~~~~$i$ &$y_{1i}$ & $y_{2i}$ & $x_{i}$  & \!$n_i$ & ~~~~$i$ &$y_{1i}$ & $y_{2i}$ & $x_i$  & $n_i$\\ 
1& \!10.18 & \!15.06 & \!0.75 & \!24 &  ~~~~10& ~8.35 & ~9.43 & \!0.47 & \!53 &~~~~19 & \!16.93 & \!16.28 & \!0.56 &~68\\
2& \!11.55 & \!17.97 & \!0.62 & \!32 &   ~~~~11& \!17.97 & \!26.82 & \!0.48 & \!56 &~~~~20 & \!11.02 & \!13.52 & \!0.34 &~68\\
3& \!16.21 & \!12.50 & \!0.66 & \!32 &   ~~~~12& \!11.84 & \!15.64 & \!0.34 & \!58 & ~~~~21 & \!14.69 & \!16.49 & \!0.56 &~72\\
4& \!12.31 & \!14.88 & \!0.26 & \!43 &   ~~~~13& \!12.43 & \!13.94 & \!0.28 & \!58 & ~~~~22 & \!10.48 & \!14.24 & \!0.79 &~77\\
5& \!12.88 & \!15.21 & \!0.96 & \!44 &   ~~~~14& \!14.73 & \!15.40 & \!0.63 & \!60 &  ~~~~23 & \!15.82 & \!15.13 & \!0.47 &~87\\
6& \!11.84 & \!17.69 & \!0.44 & \!45 &   ~~~~15 & \!15.80 & \!11.50  & \!0.26 & \!61 & ~~~~24 & \!12.66 & \!14.99 & \!0.71 & \!122\\
7& \!14.82 & \!16.91 & \!0.44 & \!48 &   ~~~~16 & \!14.81 & \!20.56 & \!0.56 & \!62 & ~~~~25 & \!10.41 & \!17.25 & \!0.45 & \!124\\
8& \!13.05 & \!15.07 & \!0.55 & \!49 &   ~~~~17 & \!11.14 & \!13.02 & \!0.02 & \!62 & ~~~~26 & \!10.32 & \!10.13 & \!0.05 & \!149\\
9& \!12.43 & \!12.01 & \!0.33 & \!51 &   ~~~~18 & \!17.12 & \!14.60 & \!0.41& \!66 & ~~~~27 & \!13.72 & \!18.18 & \!0.77 & \!198\\
\end{tabular}}\label{table2}
\end{table}

For a numerical illustration, we fit a bivariate linear mixed model on twenty\textcolor{black}{-}seven hospital profiling data that summarize whether each of 1,869 interviewed patients has a non-surgical problem or surgical one \citep{everson2000inference}. For each hospital ($i=1, 2, \ldots, 27$), the  data summarize the number of patients ($n_i$), an average of the health indices of patients as a hospital-wise severity measure ($x_i$), the percentage of non-surgical issues ($y_{1i}$) and that of surgical issues ($y_{2i}$). We tabulate the data in Table~\ref{table2}, reproducing  Table~1 of \cite{everson2000inference}. We presume that the sampling distribution of $\boldsymbol{y}^{\textrm{obs}}_i=(y_{1i}, y_{2i})^{\top}$ is approximately  a bivariate Gaussian distribution, which is also assumed by \cite{everson2000inference} because each hospital has reasonably  many patients. Since the data contain a covariate $x_i$, we fit an intercept term, i.e., $\boldsymbol{x}_i=(1, x_i)^{\top}$ for $X_i=I_p \otimes \boldsymbol{x}_i^{\top}$ in Equation~\eqref{mlmm}.   To set covariance matrices of measurement errors, \cite{everson2000inference} first calculate a common covariance matrix of $\{\boldsymbol{y}^{\textrm{obs}}_1, \boldsymbol{y}^{\textrm{obs}}_2, \ldots, \boldsymbol{y}^{\textrm{obs}}_{27}\}$, using the whole \textcolor{black}{data} of the 1,869 interviewees, and set it to $V_0$ as follows.
\begin{equation}\nonumber
V_0 = 
\left( \begin{array}{cc}
148.87 & 140.43\\
140.43 & 490.60
\end{array} \right)\!.
\end{equation}
Finally, the covariance matrix of measurement error for hospital $i$ is set to $V_i= V_0/n_i$ \textcolor{black}{so that each covariance matrix is inversely proportional to the number of patients. This results in heteroscedastic data}. 

To fit the model, we use  Gibbs samplers and EM algorithms derived under both augmentation schemes. To implement each Gibbs sampler, we draw 210,000 posterior samples of $\boldsymbol{A}$ and $\boldsymbol{\beta}$, which takes 1,314 seconds (CPU time) under DTA and 1,285 seconds under DA. The first 10,000 samples are discarded as burn-in. The initial value of each component of $\boldsymbol{\beta}$ is randomly generated from $\textrm{N}_1(0, 1)$ and that of $\boldsymbol{A}$ is randomly set to one of the known measurement error variances, $\boldsymbol{V}_i$'s. To implement the EM algorithms under both augmentation schemes, we set the initial values of $\boldsymbol{\beta}$ and $\boldsymbol{A}$ to $(0, 0, 0, 0)^{\top}$ and $I_2$, respectively.  With a tolerance level $10^{-10}$, it takes 0.927 second (183 iterations) for the DTA-based EM algorithm and 1.632 seconds (357 iterations) for the DA-based one.


\begin{figure}[b!]
\includegraphics[scale=0.33]{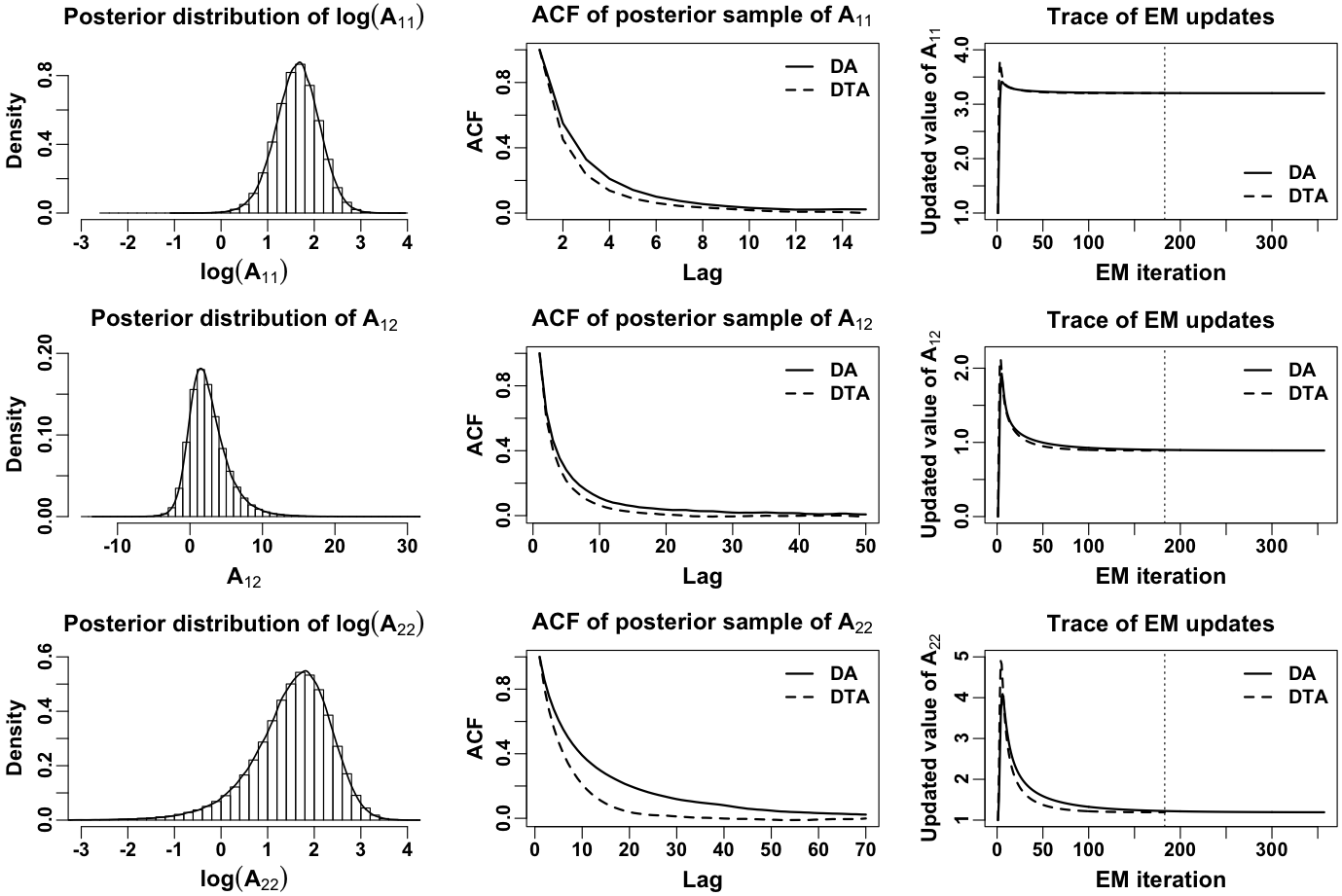}
\caption{The  results of Gibbs sampling and EM mound-climbing for (1, 1), (1, 2), and (2, 2) components of $\boldsymbol{A}$, denoted by $A_{11}, A_{12}$, and $A_{22}$, respectively. The first column shows  their posterior distributions  obtained under DTA and their  posterior densities (solid curves) obtained under DA. The second column indicates that the auto-correlation functions  under DTA (dashed curve) decrease faster than those under DA (solid curve). The last column  exhibits that the mound-climbing of the DTA-based EM algorithm (dashed curve) can be achieved with smaller number of iterations than the DA-based one (solid curve), where each vertical dotted line emphasizes the last iteration under DTA.}
\label{fig2}
\end{figure}

Figure~\ref{fig2} displays sampling and mound-climbing outcomes  for (1, 1), (1, 2), and (2, 2) components of $\boldsymbol{A}$, denoted by $A_{11}, A_{12}$, and $A_{22}$, respectively. The first column exhibits the posterior distributions of the three components obtained  under DTA. The posterior densities obtained under DA (solid curve) are superimposed to confirm that their stationary distributions  are consistent. The second column shows the auto-correlation functions of their posterior samples obtained under  DTA (dashed curve) and those under DA (solid curve). Clearly, the auto-correlation functions under DTA (dashed curve) decrease  faster for all of the three components, $A_{11}, A_{12}$, and $A_{22}$. Under DTA, the effective sample size per CPU time is 47 for $A_{11}$, 28 for $A_{12}$,  and 14 for $A_{22}$. These values are larger than their counterparts, i.e., 35 for $A_{11}$, 19 for $A_{12}$, and 7 for $A_{22}$ under DA. Each empirical \textcolor{black}{piece of} evidence  indicates the faster convergence rate of the DTA-based Gibbs sampler. The last column shows  the mode climbing result of each EM algorithm  for the three components; the vertical dotted line in each panel indicates the last iteration of the DTA-based EM algorithm. The number of iterations required under DTA is about  half of that under DA, which also contributes to the faster CPU time under DTA. 

\section{Fitting a Beta-Binomial model via DTA}\label{sec4}

A Beta-Binomial model \citep{skellam1948} assumes that the number of successes $y_i$ out of $n_i$ independent trials follows a Binomial distribution with unknown success probability $\theta_i$ (random effects), and that these random effects follow a Beta($\alpha, \beta$) distribution a priori:
\begin{equation}\label{equ4}
y^{\textrm{obs}}_i\mid \theta_i\stackrel{\textrm{ind.}}{\sim} \textrm{Bin}(n_i,~ \theta_i)~~~\textrm{and}~~~\theta_i\mid \alpha, \beta \stackrel{\textrm{i.i.d.}}{\sim} \textrm{Beta}(\alpha, \beta).
\end{equation}
A likelihood-based inference on the unknown parameters $\alpha$ and $\beta$ typically maximizes the resulting observed data likelihood function: 
\begin{equation}\label{equ5}
L(\alpha, \beta; y^{\textrm{obs}})=\prod_{i=1}^k f(y^{\textrm{obs}}_i\mid \alpha, \beta)=\prod_{i=1}^k \binom{n_i}{y^{\textrm{obs}}_i}\frac{B(y^{\textrm{obs}}_i+\alpha, ~n_i-y^{\textrm{obs}}_i+\beta)}{B(\alpha, ~\beta)},\end{equation}
where $[y^{\textrm{obs}}_i \mid \alpha, \beta]$ is an independent Beta-Binomial($\alpha, \beta$) distribution and $B(a, b)$ in \eqref{equ5} is the beta function defined as $\int_0^1u^{a-1}(1-u)^{b-1}du$. \textcolor{black}{\cite{everson2002betabin} propose a useful family of the joint prior distributions for $\alpha$ and $\beta$ whose density function is defined as $p(\alpha, \beta)\propto(\alpha+\beta+\gamma)^{-c}$. This prior is proper if $c>2$ and $\gamma>0$; for example, a popular non-informative choice with $c=2.5$ and $\gamma=0$ adopted in Chapter 5 of \cite{gelman2013bayesian} is an improper prior. The resulting posterior density is}
\begin{equation}\label{post.betabin}
p(\alpha, \beta\mid y^{\textrm{obs}})\propto L(\alpha, \beta; y^{\textrm{obs}})p(\alpha, \beta).
\end{equation}

When the number of trials is heterogeneous, i.e., $n_i$'s are not the same, it is  challenging to \textcolor{black}{analytically} integrate out one of the two parameters, $\alpha$ and $\beta$, \textcolor{black}{from $p(\alpha, \beta\mid y^{\textrm{obs}})$ in~\eqref{post.betabin}}. However, \cite{everson2002betabin} notice that if the number of trials is homogeneous ($n_i=n$) it is possible to marginalize one of the parameters from an \emph{approximate} \textcolor{black}{posterior density, $p^\ast(\alpha, \beta\mid y^{\textrm{obs}})$, defined as follows: With new notation $g(l)=nk+c+l$,}
\begin{align}
\begin{aligned}\label{equ66}
p(\alpha, \beta\mid y^{\textrm{obs}})&\approx p^\ast(\alpha, \beta\mid y^{\textrm{obs}})= \sum_{i=s_1}^{s_t}\sum_{j=f_1}^{f_t} \sum_{l=0}^{km_1+m_2}a_ib_jc^\ast_l\frac{\alpha^i\beta^j}{(\alpha+\beta+n)^{g(l)}},\\
p^\ast(\beta\mid y^{\textrm{obs}})&=\int_0^{\infty}\!\!p^\ast(\alpha, \beta\mid y^{\textrm{obs}})d\alpha=\sum_{i=s_1}^{s_t}\sum_{j=f_1}^{f_t} \sum_{l=0}^{km_1+m_2}a_ib_jc^\ast_l\frac{B(g(l)-i-1, i+1)\beta^j}{(\beta+n)^{g(l)-i-1}},
\end{aligned}
\end{align}
\textcolor{black}{where \cite{everson2002betabin} adopt an $m_1$-th order Taylor approximation at $\alpha+\beta+n$ for $L(\alpha, \beta; y^{\textrm{obs}})$ and another $m_2$-th order Taylor approximation at $\alpha+\beta+n$ for $p(\alpha, \beta)$. The notation} $s_1$ denotes the number of groups with at least one success, $s_t$ is the total number of successes ($s_t=\sum_{i=1}^ky_i^{\textrm{obs}}$), $f_1$ indicates the number of groups with at least one failure, and $f_t$ is the total number of failures ($f_t=\sum_{i=1}^k (n-y_i^{\textrm{obs}})$).  The coefficients $a_i$'s, $b_j$'s, and $c^\ast_l$'s  \textcolor{black}{are} computed via a recursive polynomial multiplication, and the function $p^\ast(\alpha, \beta\mid y^{\textrm{obs}})$ in~\eqref{equ66} can be normalized if $i+j<g(l)-2$\textcolor{black}{; see \cite{everson2002betabin} for details}.

In addition to their findings, we note that each of the  approximate marginal and conditional  density functions, $\textcolor{black}{p}^\ast(\beta\mid y^{\textrm{obs}})$ and $\textcolor{black}{p}^\ast(\alpha\mid y^{\textrm{obs}}, \beta)~(\propto \textcolor{black}{p}^\ast(\alpha, \beta\mid y^{\textrm{obs}}))$,  can be transformed into a mixture of Beta \textcolor{black}{densities}. That means, we can easily sample $\textcolor{black}{p}^\ast(\alpha, \beta\mid y^{\textrm{obs}})$ in~\eqref{equ66}  in a homoscedastic case. For example, we \textcolor{black}{can} first compute weights of \textcolor{black}{the} mixture and then sample a  Beta distribution that is randomly chosen according to the size of its weight.

This motivates the following DTA scheme:
\begin{equation}\label{equ7}
y^{\textrm{aug}}_i=h(y^{\textrm{obs}}_i, y^{\textrm{mis}}_i)=y^{\textrm{obs}}_i+y^{\textrm{mis}}_i,
\end{equation}
where $[y^{\textrm{mis}}_i\mid \theta_i]\sim\textrm{Bin}(n_\textrm{max}-n_i,~ \theta_i)$ with $n_{\textrm{max}}=\max(n_1, \ldots, n_k)$. \textcolor{black}{We set} $y^{\textrm{mis}}_i=0$ if $n_i=n_\textrm{max}$. With an assumption that $y^{\textrm{obs}}_i$ and $y^{\textrm{mis}}_i$ are conditionally independent given $\theta_i$, the augmented data become homoscedastic, i.e., $[y^{\textrm{aug}}_i\mid \theta_i]\sim\textrm{Bin}(n_\textrm{max},~ \theta_i)$. The proposed augmentation scheme iterates sampling $[y^{\textrm{mis}}\mid y^{\textrm{obs}}, \alpha, \beta]$,  $[y^{\textrm{aug}} \mid y^{\textrm{obs}},  y^{\textrm{mis}}]$ and $\textcolor{black}{p}^\ast(\alpha, \beta\mid y^{\textrm{aug}})$ in~\eqref{equ66}, where the first conditional distribution $[y^{\textrm{mis}}_i\mid y^{\textrm{obs}}_i, \alpha, \beta]$ can be easily sampled by sequentially sampling  the following two conditional distributions,
\begin{equation}\label{equ8}
[\theta_i\mid y^{\textrm{obs}}_i, \alpha, \beta]\sim\textrm{Beta}(y_i+\alpha,~ n_i-y_i+\beta)~~\textrm{and}~~[y^{\textrm{mis}}_i\mid \theta_i]\sim\textrm{Bin}(n_\textrm{max}-n_i,~ \theta_i).
\end{equation}
We specify details of this DTA scheme in the supplementary material (Section~C).


\begin{table}[t!]
\caption{\textcolor{black}{The  batting average data of ten New York Yankees baseball players during the 2019 division series versus the Minnesota Twins. The data are obtained from the Major League Baseball webpage (www.mlb.com/stats). For player $i$, the notation $n_i$ denotes the number of at-bats and $y_i$ represents the number of base hits.}}
\textcolor{black}{\begin{tabular}{ccrcccccccrcc}
$i$ & Last name  & $n_{i}$ & $y_{i}$ &~~~~~~~~$i$ & Last name & $n_{i}$ & $y_{i}$&  ~~~~~~~~$i$ & Last name & $n_{i}$ & $y_{i}$\\ 
\hline
1& Torres & 12 & 5 &  ~~~~~~~~5 & Encarnacion & 13 & 4 &~~~~~~~~8 & Urshela & 12 & 3 &\\
2& Gregorius & 10 & 4 &  ~~~~~~~~6 & LeMahieu& 14 & 4 &~~~~~~~~9 & Stanton & 6 & 1 &\\
3& Judge & 9 & 3 &  ~~~~~~~~7 & Gardner& 12 & 3 &~~~~~~~~10 & Sanchez & 8 & 1 &\\
4& Maybin & 3 & 1 &  ~~~~~ & & & &\\
\end{tabular}}\label{table3}
\end{table}

Since \cite{everson2002betabin} \textcolor{black}{do not provide a numerical illustration}, we  \textcolor{black}{check the approximation accuracy by applying the proposed DTA scheme to the batting average data of ten New York Yankees baseball players during the 2019 division series. The data are obtained from the Major League Baseball webpage and are tabulated in Table~\ref{table3}. Analyzing  over-dispersed batting average data via a hierarchical model for partially-pooled batting average estimates has been well documented  \citep{efron1975data, xianchao2012sure, tak2017rgbp}.}



\textcolor{black}{Our model-fitting configuration is as follows.} We try three different orders of the Taylor approximations for both likelihood ($m_1$) and prior ($m_2$); (i) $m=m_1=m_2=10$, (ii) $m=20$, and (iii)~$m=30$.  We first set the  prior density to $\textcolor{black}{p}(\alpha, \beta)~\textcolor{black}{\propto}~(\alpha+\beta)^{-3}$, i.e., $c=3$ and $\gamma=0$. This is a generalized Stein's harmonic prior  that is known for good frequentist coverage properties, and the resulting joint posterior distribution of $\alpha$ and $\beta$ are proper if there are at least two observations that are neither 0 nor $n_i$ \citep{tak2017propriety}.  We draw 5,100 posterior samples of $(\alpha, \beta)$ using both DTA-based Gibbs sampler  and rejection sampling. For the Gibbs sampler, initial values of $(\alpha, \beta)$ are randomly generated from the $\textrm{Gamma}(10, 1)$ distribution, and we discard the first 100 iterations as burn-in.  The effective sample size of $\alpha$ is 5,000 (out of 5,000) and that of $\beta$ is also 5,000 for all approximation orders that we try.  We use an R package \texttt{Rgbp} \citep{tak2017rgbp} to draw an exact posterior sample via a rejection sampling.

\begin{figure}[t!]
\begin{center}
\includegraphics[scale=0.43]{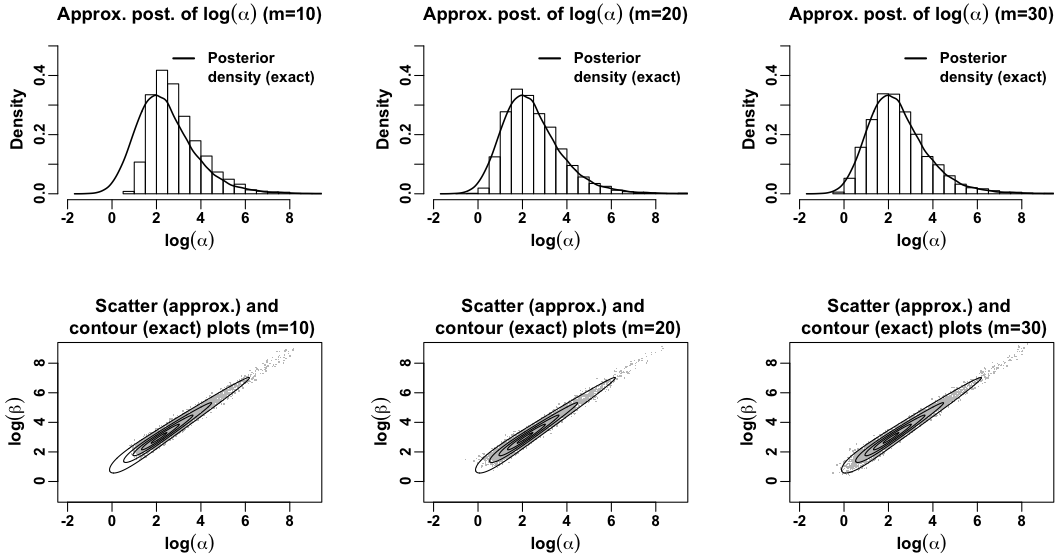}
\caption{The  results of  Gibbs sampling and rejection sampling. The panels in the first row display the posterior distributions  of $\log(\alpha)$ obtained with the \textcolor{black}{DTA}-based Gibbs sampler and superimpose the posterior densities obtained with rejection sampling (solid curves). The scatter plots in the second row are based on the approximate posterior samples of $\log(\alpha)$ and $\log(\beta)$ with contour plots (based on a grid method) superimposed. The regions with probability 5\%, 25\%, 50\%, 75\% and 95\% under the target density are outlined in the contours. As the approximation order increases, the approximate posterior distribution  obtained under the \textcolor{black}{DTA} scheme approaches the exact posterior distribution.}
\label{fig3}
\end{center}
\end{figure}



Figure~\ref{fig3} displays the sampling results. The first row shows  posterior distributions of $\log(\alpha)$ obtained with different orders of the Taylor approximation  ($m=10, 20, 30$ from left to right), and the second row exhibits  scatter plots of $\log(\alpha)$ and $\log(\beta)$ obtained under DTA  scheme with different approximation orders. For a comparison, we superimpose solid curves in the first row to display the posterior densities of $\log(\alpha)$ obtained by the rejection sampling. We also exhibit contour plots in the second row to represent the joint posterior density of $\log(\alpha)$ and $\log(\beta)$ obtained by a grid method; the contours denote the regions with probability 5\%, 25\%, 50\%, 75\% and 95\% under the target density. The posterior distributions and scatter plots obtained with DTA  clearly approach those based on the exact methods as the order increases from 10 to 30. This empirically proves the validity of the proposed DTA scheme, although it takes significant CPU time due to the recursive polynomial multiplication.


\section{Concluding remarks}\label{sec5}

Data transforming augmentation (DTA) enables transforming the observed and missing data possibly with parameters in the middle of typical  data augmentation (DA) to further improve computational simplicity and convergence rates of iterative algorithms. \textcolor{black}{To extend the applicability of DTA, we use a broader DTA framework and provide a  guideline on how to choose appropriate missing data and deterministic transformations. We also derive specific DTA schemes for converting heteroscedasticity to homoscedasticity under commonly used linear mixed and Beta-Binomial models. For the linear mixed model, we demonstrate that on average the convergence rates of the proposed DTA-based iterative algorithms are faster than those of DA-based ones, which is also empirically  confirmed  via a simulation study and realistic hospital profiling data analysis. As for the Beta-Binomial model, we test the approximation accuracy of the proposed DTA scheme, using realistic baseball data, which has not been documented in the literature.}

However, more work is expected to keep improving the  applicability of DTA. First, finding an appropriate DTA scheme  for a specific model is \textcolor{black}{still} an art, as is the case for DA  in general. \textcolor{black}{Thus, it may be better to set up a clear goal in the beginning, e.g., a goal to have the augmented data with some desired properties, and then think about what kind of missing data and transformations are needed for the desired augmented data. Since there are possibly many desired properties other than homoscedasticity, it is worthwhile to explore such vast possibilities. As for a direct extension of the current work,}  the heterogeneous covariance matrices of the observed data in linear mixed models may not be known or accurately estimable in practice, which suggests a need for a new DTA scheme for unknown covariance matrices. Also, DTA may be applicable to a linear state-space model for time series data with heteroscedastic measurement error variances. \textcolor{black}{Finally, the proposed DTA may be applicable to a Poisson-Gamma hierarchical model with heteroscedastic exposures, and its} computational advantage may be of interest.   We leave these as our future research.

\section*{Acknowledgements}\label{ack}

\textcolor{black}{Joseph Kelly and Hyungsuk Tak would like to acknowledge and thank Carl N.~Morris for the supervision and integral role played in the development of \cite{kelly2014advances}.}   Hyungsuk Tak also thanks Xiao-Li Meng for  thoughtful comments on the first draft of this manuscript and Phillip Everson for a series of productive discussions on the multivariate linear mixed model. \textcolor{black}{Finally, we thank the associate editor and the two anonymous reviewers for their insightful comments that  significantly improved the presentation of this manuscript.}



\appendix
\setcounter{secnumdepth}{0}
\section{Appendix: Comparison of matrix rates  in Section~\ref{sec31}}\label{app1}
For a $p$-dimensional parameter vector $\theta$, the matrix rate is defined as $I_p - I_{\textrm{obs}}I^{-1}_{\textrm{aug}}$, where $I_p$ is a $p$-dimensional identity matrix, 
$$
I_{\textrm{obs}}=-\frac{\partial^2 \log(L(\theta; y^{\textrm{obs}}))}{\partial \theta \partial \theta^\top}~\bigg\vert_{\theta=\hat{\theta}}
$$
is the observed Fisher information  evaluated at the maximum likelihood estimate $\hat{\theta}$, and
$$
I_{\textrm{aug}}=-E\left[\frac{\partial^2 \log(L(\theta; y^{\textrm{aug}}))}{\partial \theta \partial \theta^\top}~\bigg\vert~y^{\textrm{obs}}, \theta\right]~\bigg\vert_{\theta=\hat{\theta}}
$$
is the expected augmented information evaluated at $\hat{\theta}$, which averages over the  information from the missing data. 

For the univariate linear mixed model in~\eqref{equ1}, the observed Fisher information matrix is
$$
I_{\textrm{obs}}=
\left( \begin{array}{ccc}
\sum_{i=1}^k \frac{\boldsymbol{x}^{\top}_i\boldsymbol{x}_i}{A+V_i} &~~~~~ &\sum_{i=1}^k \frac{\boldsymbol{x}^{\top}_i(y_i-\boldsymbol{x}^{\top}_i\boldsymbol{\beta})}{(A+V_i)^2} \\
\sum_{i=1}^k \frac{\boldsymbol{x}^{\top}_i(y_i-\boldsymbol{x}^{\top}_i\boldsymbol{\beta})}{(A+V_i)^2} && -\frac{1}{2}\sum_{i=1}^k \frac{1}{(A+V_i)^2}+\sum_{i=1}^k \frac{(y_i-\boldsymbol{x}^{\top}_i\boldsymbol{\beta})^2}{(A+V_i)^3}
\end{array} \right)\bigg\vert_{(\boldsymbol{\beta}, A)=(\hat{\boldsymbol{\beta}}, \hat{A})}
$$
where the notation $(\hat{\boldsymbol{\beta}}, \hat{A})$ indicates the maximum likelihood estimates of $\boldsymbol{\beta}$ and $A$. The expected augmented data information with the  DTA scheme is 
$$
I^{\textrm{DTA}}_{\textrm{aug}}=
\left( \begin{array}{ccc}
\frac{\sum_{i=1}^k\boldsymbol{x}^{\top}_i\boldsymbol{x}_i}{A+V_{\textrm{min}}} &~~~~~ & \frac{\sum_{i=1}^k\boldsymbol{x}^{\top}_i(\mu^\ast_i-\boldsymbol{x}^{\top}_i\boldsymbol{\beta})}{(A+V_{\textrm{min}})^2} \\
  \frac{\sum_{i=1}^k\boldsymbol{x}^{\top}_i(\mu^\ast_i-\boldsymbol{x}^{\top}_i\boldsymbol{\beta})}{(A+V_{\textrm{min}})^2}  && -\frac{k}{2} \frac{1}{(A+V_{\textrm{min}})^2}+ \frac{\sum_{i=1}^k[(\mu^\ast_i-\boldsymbol{x}^{\top}_i\boldsymbol{\beta})^2+V^\ast_i]}{(A+V_{\textrm{min}})^3}
\end{array} \right)\bigg\vert_{(\boldsymbol{\beta}, A)=(\hat{\boldsymbol{\beta}}, \hat{A})}
$$
where 
\begin{align}
\mu^\ast_i&=E(y_i^{\textrm{aug}}\vert y_i^{\textrm{obs}}, \boldsymbol{\beta}, A)=(1-w_iB_i) y^{\textrm{obs}}_i+w_iB_ix^\top_i\boldsymbol{\beta},\nonumber\\
V^\ast_i&=\textrm{Var}(y_i^{\textrm{aug}}\vert y_i^{\textrm{obs}}, \boldsymbol{\beta}, A)=w_iV_{\textrm{min}}+w_i^2V_i(1-B_i).\nonumber
\end{align}
The notation $B_i$ denotes $V_i / (A+V_i)$ and $w_i$ is $1 - V_{\textrm{min}}/V_i$, as defined in Section~\ref{sec31}. The expected augmented data information with the typical DA scheme is 
$$
I^{\textrm{DA}}_{\textrm{aug}}=
\left( \begin{array}{ccc}
\frac{\sum_{i=1}^k\boldsymbol{x}^{\top}_i\boldsymbol{x}_i}{A} &~~~~~ & \frac{\sum_{i=1}^k\boldsymbol{x}^{\top}_i(\mu'_i-\boldsymbol{x}^{\top}_i\boldsymbol{\beta})}{A^2} \\
  \frac{\sum_{i=1}^k\boldsymbol{x}^{\top}_i(\mu'_i-\boldsymbol{x}^{\top}_i\boldsymbol{\beta})}{A^2}  && -\frac{k}{2A^2} + \frac{\sum_{i=1}^k[(\mu'_i-\boldsymbol{x}^{\top}_i\boldsymbol{\beta})^2+V'_i]}{A^3}
\end{array} \right)\bigg\vert_{(\boldsymbol{\beta}, A)=(\hat{\boldsymbol{\beta}}, \hat{A})}
$$
where 
\begin{align}
\mu'_i&=E(y_i^{\textrm{aug}}\vert y_i^{\textrm{obs}}, , \beta, A)=(1-B_i) y^{\textrm{obs}}_i+B_ix^\top_i\boldsymbol{\beta}\nonumber,\\
V'_i&=\textrm{Var}(y_i^{\textrm{aug}}\vert y_i^{\textrm{obs}}, \beta, A)=V_i(1-B_i).\nonumber
\end{align}
The matrix rate under DA subtracted by that under DTA is 
$$
[I_p - I_{\textrm{obs}}(I^{DA}_{\textrm{aug}})^{-1}]-[I_p - I_{\textrm{obs}}(I^{DTA}_{\textrm{aug}})^{-1}]=I_{\textrm{obs}}[(I^{DTA}_{\textrm{aug}})^{-1}-(I^{DA}_{\textrm{aug}})^{-1}]
$$
We notice that this matrix rate difference is positive definite if $I^{DA}_{\textrm{aug}}-I^{DTA}_{\textrm{aug}}$ is positive definite. However, as $I^{DA}_{\textrm{aug}}-I^{DTA}_{\textrm{aug}}$  depends on the random realization of the data $y^{\textrm{obs}}$, we take an average over $y^{\textrm{obs}}$, i.e.,
$$
E\!\left(I^{DA}_{\textrm{aug}}-I^{DTA}_{\textrm{aug}}\mid \boldsymbol{\beta}, A\right)=\left( \begin{array}{ccc}
\frac{\sum_{i=1}^k\boldsymbol{x}^{\top}_i\boldsymbol{x}_i}{A} -\frac{\sum_{i=1}^k\boldsymbol{x}^{\top}_i\boldsymbol{x}_i}{A+V_{\textrm{min}}}  &~~~~~&0 \\
0& &\frac{k}{2A^2}-\frac{k}{2(A+V_{\textrm{min}})^2}\end{array} \right)\bigg\vert_{(\boldsymbol{\beta}, A)=(\hat{\boldsymbol{\beta}}, \hat{A})}
$$
which is clearly positive definite for any non-zero minimum variance $V_{\textrm{min}}$. Thus, we can conclude that the matrix rate difference is positive definite on average, meaning that the convergence rate of the DTA-based EM algorithm is expected to be faster than the DA-based one.

\begin{appendix}
\setcounter{equation}{19}

\section{Supplementary materials}\label{sec1}

\section{A. Iterative algorithms for univariate linear mixed models in Section~3.1}

\subsection{A.1. DTA-based iterative algorithms}\label{app211}
To sample the  full posterior $p(A, \boldsymbol{\beta}\vert y^{\textrm{obs}})\propto L(A, \boldsymbol{\beta}; y^{\textrm{obs}})I_{A>0}$, \citet{kelly2014advances}  proposes the following DTA-based Gibbs-type algorithm that iteratively samples $[y^{\textrm{aug}}\vert y^{\textrm{obs}}, A, \boldsymbol{\beta}]$ and $[A, \boldsymbol{\beta}\vert y^{\textrm{aug}}]$. These two conditional distributions can be directly sampled from standard family distributions:
\begin{align}
\begin{aligned}\label{dta_lmm}
[y^{\textrm{aug}}_i\mid y^{\textrm{obs}}_i, A, \boldsymbol{\beta}] &\sim \textrm{N}_1\!\left(~(1-w_iB_i) y^{\textrm{obs}}_i+w_iB_ix^\top_i\boldsymbol{\beta},~ w_iV_{\textrm{min}}+w_i^2V_i(1-B_i)~\right),\\
[A\mid y^{\textrm{aug}}] &\sim \textrm{IG}\!\left((k-m-2)/2,~ (y^{\textrm{aug}}-X\hat{\beta})^\top(y^{\textrm{aug}}-X\hat{\beta})/2\right)~\textrm{for}~A>V_{\textrm{min}}, \\
[\boldsymbol{\beta}\mid A, y^{\textrm{aug}}] &\sim \textrm{N}_m\!\left(\hat{\boldsymbol{\beta}},~ (A+V_{\textrm{min}})(X^\top X)^{-1}\right),
\end{aligned}
\end{align}
where $w_i=1-V_{\textrm{min}}/V_i$, $B_i=V_i/(V_i+A)$, $y^{\textrm{aug}}=(y^{\textrm{aug}}_1, \ldots, y^{\textrm{aug}}_k)^\top$, IG($a, b$) denotes the inverse-Gamma distribution with shape parameter $a$ and scale parameter $b$, $X$ is a $k$ by $m$ matrix whose row vector is $\boldsymbol{x}^\top_i$, and $\hat{\boldsymbol{\beta}}=(X^\top X)^{-1}X^\top y^{\textrm{aug}}$. The second step in~\eqref{dta_lmm} can be achieved by repeatedly sampling $A$ from the inverse-Gamma distribution until $A>V_{\textrm{mis}}$ or by an inverse CDF sampling method if its cumulative distribution function and  quantile function are  available; also see Appendix of \cite{everson2000simulation}.

\citet{kelly2014advances} also derives the corresponding DTA-based EM algorithm for the posterior modes (or maximum likelihood estimates) of $A$ and $\boldsymbol{\beta}$ by constructing the following $Q$~function in the E-step: Since $[y^{\textrm{aug}}_i\vert A, \boldsymbol{\beta}] \sim \textrm{N}_1(\boldsymbol{x}^\top_i \boldsymbol{\beta}, A+V_{\min})$ under DTA,
\begin{align}\nonumber
Q(A, \beta\mid A^\ast, \boldsymbol{\beta}^\ast)&=\sum_{i=1}^kE(\log(f(y^{\textrm{aug}}_i\mid A, \boldsymbol{\beta}))\mid y^{\textrm{obs}}, A^\ast, \boldsymbol{\beta}^\ast),\nonumber\\
&=-\frac{k}{2}\log(A+V_{\min})-\frac{\sum_{i=1}^kE\left( (y^{\textrm{aug}}_i-\boldsymbol{x}^\top_i\boldsymbol{\beta})^2\mid y^{\textrm{obs}}, A^\ast, \boldsymbol{\beta}^\ast\right)}{2(A+V_{\min})}\nonumber,
\end{align}
where $A^\ast$ and $\boldsymbol{\beta}^\ast$ are the values that have maximized the $Q$ function in the previous iteration. The conditional expectation in the second equality can be computed by
\begin{equation}
E\left( (y^{\textrm{aug}}_i-\boldsymbol{x}^\top_i\boldsymbol{\beta})^2\mid y^{\textrm{obs}}, A^\ast, \boldsymbol{\beta}^\ast\right)=\left( E(y^{\textrm{aug}}_i\mid y^{\textrm{obs}}, A^\ast, \boldsymbol{\beta}^\ast)-\boldsymbol{x}^\top_i\boldsymbol{\beta}\right)^2+\textrm{Var}(y^{\textrm{aug}}_i\mid y^{\textrm{obs}}, A^\ast, \boldsymbol{\beta}^\ast)\nonumber,
\end{equation}
where the conditional mean and variance on the right-hand side, i.e., $E(y^{\textrm{aug}}_i\mid y^{\textrm{obs}}, A^\ast, \boldsymbol{\beta}^\ast)$ and $\textrm{Var}(y^{\textrm{aug}}_i\mid y^{\textrm{obs}}, A^\ast, \boldsymbol{\beta}^\ast)$, are specified in~\eqref{dta_lmm}. Maximizing this $Q$ function with respect to $\boldsymbol{\beta}$ and $A$ results in the following M-step with closed-form updates for $\boldsymbol{\beta}$ and $A$:
\begin{align}
\textrm{Step 1: }& \boldsymbol{\beta}'\gets(X^\top X)^{-1}X^\top E(y^{\textrm{aug}}\mid y^{\textrm{obs}}, A^\ast, \boldsymbol{\beta}^\ast)\nonumber,\\
\textrm{Step 2: }&A'\gets\max\left\{\frac{1}{k}\sum_{i=1}^kE\left( (y^{\textrm{aug}}_i-\boldsymbol{x}^\top_i\boldsymbol{\beta}')^2\mid y^{\textrm{obs}}, A^\ast, \boldsymbol{\beta}^\ast\right)-V_{\min},~ 0\right\}\nonumber,\\
\textrm{Step 3: }&(\boldsymbol{\beta}^\ast, A^\ast)\gets (\boldsymbol{\beta}', A')\nonumber.
\end{align}


\subsection{A.2. DA-based iterative algorithms}\label{app212}

We treat the random effects $\theta=(\theta_1, \ldots, \theta_k)^\top$ as missing data, which is typical in fitting hierarchical or mixed-effects models via DA-based iterative algorithms \citep{van2000mixed}. The resulting DA-based Gibbs algorithm  iteratively samples
\begin{align}
\begin{aligned}\label{da_lmm}
[\theta_i\mid y^{\textrm{obs}}_i, A, \boldsymbol{\beta}] &\sim \textrm{N}_1\!\left(~(1-B_i) y^{\textrm{obs}}_i+B_i\boldsymbol{x}^\top_i\boldsymbol{\beta},~ V_i(1-B_i)~\right),\\
[A\mid \theta, y^{\textrm{obs}}] & \sim \textrm{IG}\!\left((k-m-2)/2,~ (\theta-X\hat{\boldsymbol{\beta}}_{\textrm{DA}})^\top(\theta-X\hat{\boldsymbol{\beta}}_{\textrm{DA}})/2\right), \\
[\boldsymbol{\beta}\mid A, \theta, y^{\textrm{obs}}] &\sim \textrm{N}_m\!\left(\hat{\boldsymbol{\beta}}_{\textrm{DA}},~ A(X^\top X)^{-1}\right),
\end{aligned}
\end{align}
where $\hat{\boldsymbol{\beta}}_{\textrm{DA}}=(X^\top X)^{-1}X^\top \theta$. \citet{kelly2014advances} also shows a DA-based EM algorithm that corresponds to the Gibbs sampler in~\eqref{da_lmm}. Its E-step computes the $Q$ function as follows:
\begin{align}\nonumber
Q(A, \boldsymbol{\beta}\mid A^\ast, \boldsymbol{\beta}^\ast)&=\sum_{i=1}^kE(\log(f(\theta_i\mid A, \boldsymbol{\beta}))\mid y^{\textrm{obs}}, A^\ast, \boldsymbol{\beta}^\ast)\nonumber\\
&=-\frac{k}{2}\log(A)-\frac{\sum_{i=1}^kE\left( (\theta_i-\boldsymbol{x}^\top_i\boldsymbol{\beta})^2\mid y^{\textrm{obs}}, A^\ast, \boldsymbol{\beta}^\ast\right)}{2A}\nonumber,
\end{align}
where
\begin{equation}\label{da_em2}
E\left( (\theta_i-\boldsymbol{x}^\top_i\boldsymbol{\beta})^2\mid y^{\textrm{obs}}, A^\ast, \boldsymbol{\beta}^\ast\right)=\left( E(\theta_i\mid y^{\textrm{obs}}, A^\ast, \boldsymbol{\beta}^\ast)-\boldsymbol{x}^\top_i\boldsymbol{\beta}\right)^2+\textrm{Var}(\theta_i\mid y^{\textrm{obs}}, A^\ast, \boldsymbol{\beta}^\ast).
\end{equation}
The conditional mean and variance on the right-hand side of~\eqref{da_em2} are specified in~\eqref{da_lmm}. The resulting M-step sets  $\boldsymbol{\beta}^\ast$ and $A^\ast$ to the values that maximize this $Q$ function and these values are also closed-form updates as follows:
\begin{align}
\textrm{Step 1: }& \boldsymbol{\beta}'\gets(X^\top X)^{-1}X^\top E(\theta\mid y^{\textrm{obs}}, A^\ast, \boldsymbol{\beta}^\ast)\nonumber,\\
\textrm{Step 2: }&A'\gets \frac{1}{k}\sum_{i=1}^kE\left( (
\theta_i-\boldsymbol{x}^\top_i\boldsymbol{\beta})^2\mid y^{\textrm{obs}}, A^\ast, \boldsymbol{\beta}^\ast\right)\nonumber,\\
\textrm{Step 3: }&(\boldsymbol{\beta}^\ast, A^\ast)\gets (\boldsymbol{\beta}', A')\nonumber.
\end{align}

\section{B. Iterative algorithms for multivariate linear mixed models in Section~3.2}\label{app22}
\subsection{B.1. DTA-based iterative algorithms}\label{app221}

The joint posterior distribution $p(\boldsymbol{A}, \boldsymbol{\beta}\vert  \boldsymbol{y}^{\textrm{aug}})$ factors into the following two conditional distributions, $p(\boldsymbol{A}, \boldsymbol{\beta}\vert  \boldsymbol{y}^{\textrm{aug}})=p_1(\boldsymbol{A}, \vert  \boldsymbol{y}^{\textrm{aug}})p_2(\boldsymbol{\beta}\vert \boldsymbol{A},  \boldsymbol{y}^{\textrm{aug}})$, and these can be directly sampled via inverse-Wishart and multivariate Gaussian distributions in a homoscedastic case. A DTA-based Gibbs-type algorithm specified in~(3) iteratively samples the following three conditional distributions.
\begin{align}
\begin{aligned}\label{dta_mlmm}
[\boldsymbol{y}^{\textrm{aug}}_i\mid \boldsymbol{y}^{\textrm{obs}}_i, \boldsymbol{A}, \boldsymbol{\beta}] &\sim \textrm{N}_p\!\left((1-W_iB_i) \boldsymbol{y}^{\textrm{obs}}_i+W_iB_iX_i\boldsymbol{\beta},~ \boldsymbol{V}\!_{\textrm{min}}W_i^\top+W_i(1-B_i)\boldsymbol{V}\!_iW_i^\top\right),\\
[\boldsymbol{A}+\boldsymbol{V}\!_{\textrm{min}}\mid \boldsymbol{y}^{\textrm{aug}}] & \sim \textrm{IW}\!\left(k-m-p-1,~ \sum_{i=1}^k(\boldsymbol{y}^{\textrm{aug}}_i-X_i\hat{\boldsymbol{\beta}}_{\textrm{DTA}})(\boldsymbol{y}^{\textrm{aug}}_i-X_i\hat{\boldsymbol{\beta}}_{\textrm{DTA}})^\top\right),\\
[\boldsymbol{\beta}\mid \boldsymbol{A}, \boldsymbol{y}^{\textrm{aug}}] &\sim \textrm{N}_{mp}\!\left(\hat{\boldsymbol{\beta}}_{\textrm{DTA}},~ \left(\sum_{i=1}^kX^\top_i (\boldsymbol{A}+\boldsymbol{V}\!_{\textrm{min}})^{-1}X_i\right)^{-1}\right),
\end{aligned}
\end{align}
where $B_i=\boldsymbol{V}\!_i(\boldsymbol{V}\!_i+\boldsymbol{A})^{-1}$, IW($a, b$) indicates the inverse-Wishart distribution with $a$ degrees of freedom and scale matrix $b$, and
\begin{equation}\nonumber
\hat{\boldsymbol{\beta}}_{\textrm{DTA}}= \left(\sum_{i=1}^kX^\top_i (\boldsymbol{A}+\boldsymbol{V}\!_{\textrm{min}})^{-1}X_i\right)^{-1}\sum_{i=1}^kX^\top_i(\boldsymbol{A}+\boldsymbol{V}\!_{\textrm{min}})^{-1} \boldsymbol{y}^{\textrm{aug}}_i.
\end{equation}
To  sample $\boldsymbol{A}$ instead of $\boldsymbol{A}+\boldsymbol{V}\!_{\textrm{min}}$ in the middle of~\eqref{dta_mlmm}, we   repeatedly draw a random sample $K$ from the inverse-Wishart distribution in~\eqref{dta_mlmm} until $\vert K- \boldsymbol{V}\!_{\textrm{min}}\vert>0$, and then  set $\boldsymbol{A}$ to $K- \boldsymbol{V}\!_{\textrm{min}}$.

We specify the corresponding DTA-based EM algorithm  by constructing the $Q$ function for the E-step, using the marginal distribution $[\boldsymbol{y}_i^{\textrm{aug}}\vert \boldsymbol{A}, \boldsymbol{\beta}]  \sim \textrm{N}_p(\boldsymbol{X}\!_i\boldsymbol{\beta},~ \boldsymbol{A}+\boldsymbol{V}\!_{\textrm{min}})$:
\begin{align}
\begin{aligned}\label{dta_m}
&Q(\boldsymbol{A}, \boldsymbol{\beta}\mid \boldsymbol{A}^\ast, \boldsymbol{\beta}^\ast)=\sum_{i=1}^kE\left(\log(f(\boldsymbol{y}^{\textrm{aug}}_i\mid \boldsymbol{A}, \boldsymbol{\beta}))\mid \boldsymbol{y}^{\textrm{obs}}, \boldsymbol{A}^\ast, \boldsymbol{\beta}^\ast\right)\\
&=-\frac{k}{2}\log(\vert \boldsymbol{A}+\boldsymbol{V}\!_{\min}\vert)-\frac{1}{2}\sum_{i=1}^kE\left( (\boldsymbol{y}^{\textrm{aug}}_i-X_i\boldsymbol{\beta})^\top(\boldsymbol{A}+\boldsymbol{V}\!_{\textrm{min}})^{-1}(\boldsymbol{y}^{\textrm{aug}}_i-X_i\boldsymbol{\beta})\mid \boldsymbol{y}^{\textrm{obs}}, \boldsymbol{A}^\ast, \boldsymbol{\beta}^\ast\right).
\end{aligned}
\end{align}
The conditional expectation of a quadratic form in the second equality is equivalent to
\begin{align}
\begin{aligned}\label{dta_em_mlmm}
E&\left( (\boldsymbol{y}^{\textrm{aug}}_i-X_i\boldsymbol{\beta})^\top(\boldsymbol{A}+\boldsymbol{V}\!_{\textrm{min}})^{-1}(\boldsymbol{y}^{\textrm{aug}}_i-X_i\boldsymbol{\beta})\mid \boldsymbol{y}^{\textrm{obs}}, \boldsymbol{A}^\ast, \boldsymbol{\beta}^\ast\right)\\
&=\left( E(\boldsymbol{y}^{\textrm{aug}}_i \mid \boldsymbol{y}^{\textrm{obs}}, \boldsymbol{A}^\ast, \boldsymbol{\beta}^\ast)-X_i\boldsymbol{\beta}\right)^\top(\boldsymbol{A}+\boldsymbol{V}\!_{\textrm{min}})^{-1}\left( E(\boldsymbol{y}^{\textrm{aug}}_i\mid \boldsymbol{y}^{\textrm{obs}}, \boldsymbol{A}^\ast, \boldsymbol{\beta}^\ast)-X_i\boldsymbol{\beta}\right)\\
&\textcolor{white}{==}+\textrm{trace}\!\left[ (\boldsymbol{A}+\boldsymbol{V}\!_{\textrm{min}})^{-1}\textrm{Cov}(\boldsymbol{y}^{\textrm{aug}}_i\mid \boldsymbol{y}^{\textrm{obs}}, \boldsymbol{A}^\ast, \boldsymbol{\beta}^\ast)\right].
\end{aligned}
\end{align}
The conditional expectation and covariance of $\boldsymbol{y}^{\textrm{aug}}_i$ on the right-hand side in~\eqref{dta_em_mlmm}, i.e., $E(\boldsymbol{y}^{\textrm{aug}}_i \mid \boldsymbol{y}^{\textrm{obs}}, \boldsymbol{A}^\ast, \boldsymbol{\beta}^\ast)$ and $\textrm{Cov}(\boldsymbol{y}^{\textrm{aug}}_i\mid \boldsymbol{y}^{\textrm{obs}}, \boldsymbol{A}^\ast, \boldsymbol{\beta}^\ast)$, are specified in~\eqref{dta_mlmm}.

The M-step updates $\boldsymbol{A}^\ast$ and $\boldsymbol{\beta}^\ast$ by the values that maximize the $Q$ function in~\eqref{dta_m}, which results in the following four steps for closed-form updates:
\begin{align}
\begin{aligned}
\textrm{Step 1: }& \boldsymbol{\beta}' \gets\left(\sum_{i=1}^kX_i^\top X_i\right)^{-1}\sum_{i=1}^kX_i^\top E(
\boldsymbol{y}^{\textrm{aug}}_i\mid \boldsymbol{y}^{\textrm{obs}}, \boldsymbol{A}^\ast, \boldsymbol{\beta}^\ast).\nonumber\\
\textrm{Step 2: }& \boldsymbol{A}_{\textrm{temp}}  \gets \frac{1}{k}\sum_{i=1}^k\bigg\{\left(E(\boldsymbol{y}^{\textrm{aug}}_i \mid \boldsymbol{y}^{\textrm{obs}}, \boldsymbol{A}^\ast, \boldsymbol{\beta}^\ast)-X_i\boldsymbol{\beta}'\right)\left(E(\boldsymbol{y}^{\textrm{aug}}_i\mid \boldsymbol{y}^{\textrm{obs}}, \boldsymbol{A}^\ast, \boldsymbol{\beta}^\ast)-X_i\boldsymbol{\beta}'\right)^\top\\
&\textcolor{white}{==========}+\textrm{Cov}\left(\boldsymbol{y}^{\textrm{aug}}_i\mid \boldsymbol{y}^{\textrm{obs}}, \boldsymbol{A}^\ast, \boldsymbol{\beta}^\ast\right)\bigg\}-\boldsymbol{V}\!_{\textrm{min}}.\\
\textrm{Step 3: }& \boldsymbol{A}'  \gets \boldsymbol{A}_{\textrm{temp}}~~\textrm{if}~~\vert \boldsymbol{A}_{\textrm{temp}}\vert>0~~ \textrm{and}~~ \boldsymbol{A}^\ast  \gets 0_p~~\textrm{otherwise}.\\
\textrm{Step 4: }& (\boldsymbol{\beta}^\ast, \boldsymbol{A}^\ast)  \gets (\boldsymbol{\beta}', \boldsymbol{A}').
\end{aligned}
\end{align}
The notation $0_p$ in Step~3 indicates a $p$ by $p$ matrix filled with zeros.

\subsection{B.2. DA-based iterative algorithms}\label{app222}

In a typical DA scheme, we treat the random effects, $\boldsymbol{\theta}\equiv\{\boldsymbol{\theta}_1, \ldots, \boldsymbol{\theta}_k\}$, as missing data, and thus the augmented data in this case are $\boldsymbol{y}^{\textrm{aug}}=(\boldsymbol{y}^{\textrm{obs}}, \boldsymbol{\theta})$. The full posterior density function of $[\boldsymbol{\theta},  \boldsymbol{\beta}, \boldsymbol{A}\vert \boldsymbol{y}^{\textrm{obs}}]$ can be derived up to a constant multiplication, i.e.,
$$
\pi(\boldsymbol{\theta},  \boldsymbol{\beta}, \boldsymbol{A}\mid \boldsymbol{y}^{\textrm{obs}})\propto \prod_{i=1}^kf(\boldsymbol{y}^{\textrm{obs}}_i\mid \boldsymbol{\theta}_i)p(\boldsymbol{\theta}_i\mid  \boldsymbol{\beta}, \boldsymbol{A})I_{\vert \boldsymbol{A}\vert>0},
$$
where $[\boldsymbol{y}^{\textrm{obs}}_i\vert \boldsymbol{\theta}_i]$ and $[\boldsymbol{\theta}_i\vert \boldsymbol{A}, \boldsymbol{\beta}]$ are defined in~(11). Similarly to the DTA-based Gibbs sampler in~\eqref{dta_mlmm},  the DA-based one iteratively samples the following conditional distributions:
\begin{align}
\begin{aligned}\label{da_mlmm}
[\boldsymbol{\theta}_i\mid \boldsymbol{y}^{\textrm{obs}}_i, \boldsymbol{A}, \boldsymbol{\beta}] &\sim \textrm{N}_p\!\left(~(1-B_i) \boldsymbol{y}^{\textrm{obs}}_i+B_iX_i\boldsymbol{\beta},~ (1-B_i)\boldsymbol{V}\!_i~\right),\\
[\boldsymbol{A}\mid \boldsymbol{\theta}, \boldsymbol{y}^{\textrm{obs}}] & \sim \textrm{IW}\!\left(~k-m-p-1,~ \sum_{i=1}^k(\boldsymbol{\theta}_i-X_i\hat{\boldsymbol{\beta}}_{\textrm{DA}})(\boldsymbol{\theta}_i-X_i\hat{\boldsymbol{\beta}}_{\textrm{DA}})^\top\right), \\
[\boldsymbol{\beta}\mid \boldsymbol{A}, \boldsymbol{\theta}, \boldsymbol{y}^{\textrm{obs}}] &\sim \textrm{N}_{mp}\!\left(\hat{\boldsymbol{\beta}}_{\textrm{DA}},~ \left(\sum_{i=1}^kX^\top_i \boldsymbol{A}^{-1}X_i\right)^{-1}\right),
\end{aligned}
\end{align}
where $B_i=\boldsymbol{V}\!_i(\boldsymbol{V}\!_i+\boldsymbol{A})^{-1}$ and
$$
\hat{\boldsymbol{\beta}}_{\textrm{DA}}= \left(\sum_{i=1}^kX^\top_i \boldsymbol{A}^{-1}X_i\right)^{-1}\sum_{i=1}^kX^\top_i \boldsymbol{A}^{-1}\boldsymbol{\theta}_i.
$$

The corresponding DA-based EM algorithm adopts the following $Q$ function for the E-step, using the distribution of missing data, $[\boldsymbol{\theta}_i\vert \boldsymbol{A}, \boldsymbol{\beta}]  \sim \textrm{N}_p(\boldsymbol{X}\!_i\boldsymbol{\beta},~ \boldsymbol{A})$:
\begin{align}
\begin{aligned}\label{da_m}
Q(\boldsymbol{A}, \boldsymbol{\beta}\mid &\boldsymbol{A}^\ast, \boldsymbol{\beta}^\ast)=\sum_{i=1}^kE(\log(f(\boldsymbol{\theta}_i\mid \boldsymbol{A}, \boldsymbol{\beta}))\mid \boldsymbol{y}^{\textrm{obs}}, \boldsymbol{A}^\ast, \boldsymbol{\beta}^\ast),\\
&=-\frac{k}{2}\log(\vert \boldsymbol{A}\vert)-\frac{1}{2}\sum_{i=1}^kE\left( (\boldsymbol{\theta}_i-X_i\boldsymbol{\beta})^\top \boldsymbol{A}^{-1}(\boldsymbol{\theta}_i-X_i\boldsymbol{\beta})\mid \boldsymbol{y}^{\textrm{obs}}, \boldsymbol{A}^\ast, \boldsymbol{\beta}^\ast\right).
\end{aligned}
\end{align}
The conditional expectation of a quadratic form on the right-hand side in~\eqref{da_m} can be computed by
\begin{align}
\begin{aligned}\label{da_em_mlmm}
E&\left( (\boldsymbol{\theta}_i-X_i\boldsymbol{\beta})^\top A^{-1}(\boldsymbol{\theta}_i-X_i\boldsymbol{\beta})\mid \boldsymbol{y}^{\textrm{obs}}, \boldsymbol{A}^\ast, \boldsymbol{\beta}^\ast\right)\\
&=\left( E(\boldsymbol{\theta}_i \mid \boldsymbol{y}^{\textrm{obs}}, \boldsymbol{A}^\ast, \boldsymbol{\beta}^\ast)-X_i\boldsymbol{\beta}\right)^\top \boldsymbol{A}^{-1}\left( E(\boldsymbol{\theta}_i\mid \boldsymbol{y}^{\textrm{obs}}, \boldsymbol{A}^\ast, \boldsymbol{\beta}^\ast)-X_i\boldsymbol{\beta}\right)\\
&\textcolor{white}{==}+\textrm{trace}\!\left[ \boldsymbol{A}^{-1}\textrm{Cov}(\boldsymbol{\theta}_i\mid \boldsymbol{y}^{\textrm{obs}}, \boldsymbol{A}^\ast, \boldsymbol{\beta}^\ast)\right].
\end{aligned}
\end{align}
The conditional expectation and covariance of $\boldsymbol{\theta}_i$ given $\boldsymbol{y}^{\textrm{obs}}, \boldsymbol{A}, \boldsymbol{\beta}$ in~\eqref{da_em_mlmm} are specified in~\eqref{da_mlmm}.

Like the M-step under DTA in~\eqref{dta_m}, the M-step under DA updates $\boldsymbol{A}^\ast$ and $\boldsymbol{\beta}^\ast$ by the values that maximize the $Q$ function in~\eqref{da_m} via the following three steps:
\begin{align}
\begin{aligned}
\textrm{Step 1: }& \boldsymbol{\beta}' \gets\left(\sum_{i=1}^kX_i^\top X_i\right)^{-1}\sum_{i=1}^kX_i^\top E(
\boldsymbol{\theta}_i\mid \boldsymbol{y}^{\textrm{obs}}, \boldsymbol{A}^\ast, \boldsymbol{\beta}^\ast).\nonumber\\
\textrm{Step 2: }& \boldsymbol{A}' \gets \frac{1}{k}\sum_{i=1}^k\bigg\{\left(E(\boldsymbol{\theta}_i \mid \boldsymbol{y}^{\textrm{obs}}, \boldsymbol{A}^\ast, \beta^\ast)-X_i\boldsymbol{\beta}'\right)\left(E(\boldsymbol{\theta}_i\mid \boldsymbol{y}^{\textrm{obs}}, \boldsymbol{A}^\ast, \boldsymbol{\beta}^\ast)-X_i\boldsymbol{\beta}'\right)^\top\\
&\textcolor{white}{========}+\textrm{Cov}\!\left(\boldsymbol{\theta}_i\mid \boldsymbol{y}^{\textrm{obs}}, \boldsymbol{A}^\ast, \boldsymbol{\beta}^\ast\right)\bigg\}.\\
\textrm{Step 3: }& (\boldsymbol{\beta}^\ast, \boldsymbol{A}^\ast) \gets (\boldsymbol{\beta}', \boldsymbol{A}').\\
\end{aligned}
\end{align}

\section{C. The DTA scheme for the Beta-Binomial model  in Section~4}\label{app4}

Given the homoscedastic augmented data $y^{\textrm{aug}}$, we  reproduce the approximate marginal posterior density $p^\ast(\beta\mid y^{\textrm{aug}})$ from~(17): With $g(l)=nk+c+l$, 
$$
p^\ast(\beta\mid y^{\textrm{aug}})=\sum_{i=s_1}^{s_t}\sum_{j=f_1}^{f_t} \sum_{l=0}^{km_1+m_2}a_ib_jc^\ast_lB(g(l)-i-1,~ i+1)\frac{\beta^j}{(\beta+n_{\textrm{max}})^{g(l)-i-1}},
$$
where $s_1$ denotes the number of groups with at least one success, $s_t$ is the total number of successes ($s_t=\sum_{i=1}^ky_i^{\textrm{aug}}$), $f_1$ indicates the number of groups with at least one failure, and $f_t$ is the total number of failures ($f_t=\sum_{i=1}^k (n-y_i^{\textrm{aug}})$). If we transform $\beta$ into $B=\beta/(\beta+n_{\textrm{max}})$, the corresponding density function with respect to $B$ is as follows: With the Jacobean $ J =n_{\textrm{max}}/(1-B)^2$,
$$
p^\ast(B\mid y^{\textrm{aug}})=\sum_{i=s_1}^{s_t}\sum_{j=f_1}^{f_t} \sum_{l=0}^{km_1+m_2}a_ib_jc^\ast_lB(g(l)-i-1,~ i+1)\frac{B^j(1-B)^{g(l)-i-j-3}}{n^{g(l)-i-j-1}_{\textrm{max}}}.
$$
Thus, each mixture component is composed of the Beta$(j+1,~ g(l)-i-j-2)$ density function  and the corresponding coefficient (that is proportional to its weight) equal to 
\begin{equation}\label{weight1}
a_ib_jc^\ast_lB(g(l)-i-1,~ i+1)B(j+1,~ g(l)-i-j-2)n^{-(g(l)-i-j-1)}_{\textrm{max}}.
\end{equation}
We can easily generate $B=\beta/(\beta+n_{\textrm{max}})$ via a two-step procedure; (i) we randomly choose a combination of $(i, j, l)$ according to its weight defined in~\eqref{weight1}, and (ii) given the selected values of $(i, j, l)$, we can generate $B$ from the $\textrm{Beta}(j+1,~ g(l)-i-j-2)$ distribution. Finally, we set $\beta=n_{\textrm{max}}B/(1-B)$ that is a random number generated from $p^\ast(\beta\vert y^{\textrm{aug}})$.

Given the random number from $p^\ast(\beta\vert y^{\textrm{aug}})$, we need to sample  $p^\ast(\alpha\vert y^{\textrm{aug}}, \beta)$ that is proportional to $p^\ast(\alpha, \beta \vert y^{\textrm{aug}})$ in~(16), i.e., 
$$
p^\ast(\alpha\mid y^{\textrm{aug}}, \beta)\propto\sum_{i=s_1}^{s_t}\sum_{j=f_1}^{f_t} \sum_{l=0}^{km_1+m_2}a_ib_jc^\ast_l\frac{\alpha^i\beta^j}{(\alpha+\beta+n_{\textrm{max}})^{g(l)}}.
$$
Once we transform $\alpha$ to $A=\alpha/(\alpha+\beta+n_{\textrm{max}})$, we obtain the following density function with the Jacobean $J=(\beta+n_{\textrm{max}}) / (1-A)^2$:
\begin{align}
\begin{aligned}
p^\ast(A\mid y^{\textrm{aug}}, \beta)&\propto\sum_{i=s_1}^{s_t}\sum_{j=f_1}^{f_t}\sum_{l=0}^{km_1+m_2}a_ib_jc^\ast_l\frac{\beta^j}{(\beta+n_{\textrm{max}})^{g(l)-i-1}}A^i(1-A)^{g(l)-i-2}\\
&=\left(\sum_{j=f_1}^{f_t}b_j\beta^j\right)\left(\sum_{i=s_1}^{s_t}\sum_{l=0}^{km_1+m_2}a_ic^\ast_l\frac{1}{(\beta+n_{\textrm{max}})^{g(l)-i-1}}A^i(1-A)^{g(l)-i-2}\right)\\
&\propto\sum_{i=s_1}^{s_t}\sum_{l=0}^{km_1+m_2}a_ic^\ast_l\frac{1}{(\beta+n_{\textrm{max}})^{l-i-1}}A^i(1-A)^{g(l)-i-2}.
\end{aligned}
\end{align}
Then, the density with respect to $A$ is a mixture of the Beta$(i+1,~ g(l)-i-1)$ densities with its coefficient (weight) equal to
\begin{equation}\label{weight2}
a_ic^\ast_l\frac{B(i+1,~ g(l)-i-1)}{(\beta+n_{\textrm{max}})^{l-i-1}}.
\end{equation}
Sampling $\alpha$ from $p^\ast(\alpha\mid y^{\textrm{aug}}, \beta)$ is a three-step procedure; (i) a combination of $(i, l)$ is randomly selected according to its weight in~\eqref{weight2},  (ii) a value of $A$ is randomly generated from the Beta$(i+1,~ g(l)-i-1)$ distribution given the chosen values of $(i, l)$, and finally (iii) $\alpha$ is set to $(n_{\textrm{max}}+\beta)A/(1-A)$.

Therefore, the proposed augmentation scheme for heteroscedastic Binomial data  results in a Gibbs-type algorithm that iterates for following five steps:
\begin{align}
\begin{aligned}\nonumber
\textrm{Step 1: }& \textrm{Sample } \theta_i\sim\textrm{Beta}(y_i^{\textrm{obs}}+\alpha,~ n_i-y_i^{\textrm{obs}}+\beta)~\textrm{for }i=1, \ldots, k.\\
\textrm{Step 2: }& \textrm{Sample } y^{\textrm{mis}}_i\sim\textrm{Bin}(n_{\textrm{max}}-n_i,~ \theta_i)~\textrm{for }i=1, \ldots, k.\\
\textrm{Step 3: }& \textrm{Set } y^{\textrm{aug}}_i=y^{\textrm{obs}}_i+y^{\textrm{mis}}_i~\textrm{for }i=1, \ldots, k.\\
\textrm{Step 4: }& \textrm{Sample } \beta~ \textrm{from } p^\ast(\beta\mid y^{\textrm{aug}}).\\
\textrm{Step 5: }& \textrm{Sample } \alpha~ \textrm{from } p^\ast(\alpha\mid y^{\textrm{aug}}, \beta).
\end{aligned}
\end{align}
We note that Steps 1--3 are corresponding to the first two steps of~(2), and Steps 4--5 are related to the last step of~(2).

\end{appendix}

\bibliographystyle{apalike}

\bibliography{bibliography}

\begin{thebibliography}{}

\bibitem[Daniels, 1999]{daniels1999prior}
Daniels, M.~J. (1999).
\newblock {A Prior for the Variance in Hierarchical Models}.
\newblock {\em Canadian Journal of Statistics}, 27(3):567--578.

\bibitem[Dempster et~al., 1977]{dempster1977maximum}
Dempster, A.~P., Laird, N.~M., and Rubin, D.~B. (1977).
\newblock {Maximum Likelihood from Incomplete Data via the EM Algorithm}.
\newblock {\em Journal of the Royal Statistical Society.~Series B},
  39(1):1--38.

\bibitem[Efron and Morris, 1975]{efron1975data}
Efron, B. and Morris, C.~N. (1975).
\newblock {Data Analysis Using Stein's Estimator and its Generalizations}.
\newblock {\em Journal of the American Statistical Association},
  70(350):311--319.

\bibitem[Everson and Bradlow, 2002]{everson2002betabin}
Everson, P.~J. and Bradlow, E.~T. (2002).
\newblock {Bayesian Inference for the Beta-Binomial Distribution via Polynomial
  Expansions}.
\newblock {\em Journal of Computational and Graphical Statistics},
  11(1):202--207.

\bibitem[Everson and Morris, 2000a]{everson2000inference}
Everson, P.~J. and Morris, C.~N. (2000a).
\newblock {Inference for Multivariate Normal Hierarchical Models}.
\newblock {\em Journal of the Royal Statistical Society.~Series B},
  62(2):399--412.

\bibitem[Everson and Morris, 2000b]{everson2000simulation}
Everson, P.~J. and Morris, C.~N. (2000b).
\newblock {Simulation from Wishart Distributions with Eigenvalue Constraints}.
\newblock {\em Journal of Computational and Graphical Statistics},
  9(2):380--389.

\bibitem[Gasparrini et~al., 2012]{gasparrini2012mlmm}
Gasparrini, A., Armstrong, B., and Kenward, M.~G. (2012).
\newblock {Multivariate Meta-Analysis for Non-Linear and Other Multi-Parameter
  Associations}.
\newblock {\em Statistics in Medicine}, 31(29):3821--3839.

\bibitem[Gelman et~al., 2013]{gelman2013bayesian}
Gelman, A., Carlin, J.~B., Stern, H.~S., Dunson, D.~B., Vehtari, A., and Rubin,
  D.~B. (2013).
\newblock {\em {Bayesian Data Analysis}}.
\newblock CRC Press, Boca Raton, FL, USA.

\bibitem[Geman and Geman, 1984]{geman1984stochastic}
Geman, S. and Geman, D. (1984).
\newblock {Stochastic Relaxation, Gibbs Distributions, and the Bayesian
  Restoration of Images}.
\newblock {\em IEEE Transactions on Pattern Analysis and Machine Intelligence},
  PAMI-6(6):721--741.

\bibitem[Kass and Steffey, 1989]{kass1989approx}
Kass, R.~E. and Steffey, D. (1989).
\newblock {Approximate Bayesian Inference in Conditionally Independent
  Hierarchical Models (Parametric Empirical Bayes Models)}.
\newblock {\em Journal of the American Statistical Association},
  84(407):717--726.

\bibitem[Kelly, 2014]{kelly2014advances}
Kelly, J. (2014).
\newblock {\em {Advances in the Normal-Normal Hierarchical Model}}.
\newblock PhD thesis, Harvard University.

\bibitem[Liu et~al., 1998]{liu1998pxem}
Liu, C., Rubin, D.~B., and Wu, Y.~N. (1998).
\newblock {Parameter Expansion to Accelerate EM: The PX-EM Algorithm}.
\newblock {\em Biometrika}, 85(4):755--770.

\bibitem[Meng and van Dyk, 1997]{meng1997folksong}
Meng, X.-L. and van Dyk, D.~A. (1997).
\newblock {The EM Algorithm--An Old Folk-Song Sung to a Fast New Tune}.
\newblock {\em Journal of the Royal Statistical Society.~Series B},
  59(3):511--567.

\bibitem[Meng and van Dyk, 1999]{meng1999seeking}
Meng, X.-L. and van Dyk, D.~A. (1999).
\newblock {Seeking Efficient Data Augmentation Schemes via Conditional and
  Marginal Augmentation}.
\newblock {\em Biometrika}, 86(2):301--320.

\bibitem[Morris, 1987]{morris1987comment}
Morris, C.~N. (1987).
\newblock {The Calculation of Posterior Distributions by Data Augmentation:
  Comment: Simulation in Hierarchical Models}.
\newblock {\em Journal of the American Statistical Association},
  82(398):542--543.

\bibitem[Morris and Lysy, 2012]{morris2012shrinkage}
Morris, C.~N. and Lysy, M. (2012).
\newblock {Shrinkage Estimation in Multilevel Normal Models}.
\newblock {\em Statistical Science}, 27(1):115--134.

\bibitem[Papaspiliopoulos and Roberts, 2008]{papa2008repara}
Papaspiliopoulos, O. and Roberts, G.~O. (2008).
\newblock {Stability of the Gibbs Sampler for Bayesian Hierarchical Models}.
\newblock {\em The Annals of Statistics}, 36(1):95--117.

\bibitem[Papaspiliopoulos et~al., 2007]{papa2007repara}
Papaspiliopoulos, O., Roberts, G.~O., and Sk\''old, M. (2007).
\newblock {A General Framework for the Parametrization of Hierarchical Models}.
\newblock {\em Statistical Science}, 22(1):59--73.

\bibitem[{R Development Core Team}, 2019]{r2019}
{R Development Core Team} (2019).
\newblock {\em {R: A Language and Environment for Statistical Computing}}.
\newblock R Foundation for Statistical Computing, Vienna, Austria.

\bibitem[Skellam, 1948]{skellam1948}
Skellam, J.~G. (1948).
\newblock {A Probability Distribution Derived from the Binomial Distribution by
  Regarding the Probability of Success as Variable Between the Sets of Trials}.
\newblock {\em Journal of the Royal Statistical Society.~Series B},
  10(2):257--261.

\bibitem[Staudenmayer et~al., 2008]{staudenmayer2008density}
Staudenmayer, J., Ruppert, D., and Buonaccorsi, J.~P. (2008).
\newblock {Density Estimation in the Presence of Heteroscedastic Measurement
  Error}.
\newblock {\em Journal of the American Statistical Association},
  103(482):726--736.

\bibitem[Tak, 2017]{tak2017frequency}
Tak, H. (2017).
\newblock {Frequency Coverage Properties of a Uniform Shrinkage Prior
  Distribution}.
\newblock {\em Journal of Statistical Computation and Simulation},
  87(15):2929--2939.

\bibitem[Tak et~al., 2017]{tak2017rgbp}
Tak, H., Kelly, J., and Morris, C.~N. (2017).
\newblock {Rgbp: An R Package for Gaussian, Poisson, and Binomial Random
  Effects Models, with Frequency Coverage Evaluations}.
\newblock {\em Journal of Statistical Software}, 78(5):1--33.

\bibitem[Tak and Morris, 2017]{tak2017propriety}
Tak, H. and Morris, C.~N. (2017).
\newblock {Data-Dependent Posterior Propriety of a Bayesian Beta-Binomial-Logit
  Model}.
\newblock {\em Bayesian Analysis}, 12(2):533--555.

\bibitem[Tanner and Wong, 1987]{tanner1987calculation}
Tanner, M.~A. and Wong, W.~H. (1987).
\newblock {The Calculation of Posterior Distributions by Data Augmentation}.
\newblock {\em Journal of the American Statistical Association},
  82(398):528--540.

\bibitem[van Dyk, 2000]{van2000mixed}
van Dyk, D.~A. (2000).
\newblock {Fitting Mixed-Effects Models Using Efficient EM-Type Algorithms}.
\newblock {\em Journal of Computational and Graphical Statistics}, 9(1):78--98.

\bibitem[van Dyk and Meng, 2001]{van2001art}
van Dyk, D.~A. and Meng, X.-L. (2001).
\newblock {The Art of Data Augmentation}.
\newblock {\em Journal of Computational and Graphical Statistics}, 10(1):1--50.

\bibitem[van Dyk and Meng, 2010]{vandyk2010fertilizing}
van Dyk, D.~A. and Meng, X.-L. (2010).
\newblock {Cross-Fertilizing Strategies for Better EM Mountain Climbing and DA
  Field Exploration: A Graphical Guide Book}.
\newblock {\em Statistical Science}, 25(4):429--449.

\bibitem[Xie et~al., 2012]{xianchao2012sure}
Xie, X., Kou, S.~C., and Brown, L.~D. (2012).
\newblock {SURE Estimates for a Heteroscedastic Hierarchical Model}.
\newblock {\em Journal of the American Statistical Association},
  107(500):1465--1479.

\bibitem[Yu and Meng, 2011]{yu2011}
Yu, Y. and Meng, X.-L. (2011).
\newblock {To Center or not to Center: That is not the Question? An
  Ancillarity--Sufficiency Interweaving Strategy (ASIS) for Boosting MCMC
  Efficiency}.
\newblock {\em Journal of Computational and Graphical Statistics},
  20(3):531--570.

\end{thebibliography}

\end{document}